\documentclass[useAMS,usenatbib]{mn2e}
\usepackage[colorlinks=true,allcolors=blue]{hyperref}

\usepackage{natbib}
\usepackage{color,graphicx} 
\usepackage{amsmath}        
\usepackage{amsfonts}       
\usepackage{amssymb}        
\usepackage{gensymb}        
\usepackage{framed}         
\usepackage{upgreek}
\usepackage{xspace}
\usepackage{wrapfig}
\usepackage{rotating}
\usepackage[space]{grffile}
\usepackage{latexsym}
\usepackage{textcomp}
\usepackage{longtable}
\usepackage{multirow,booktabs}
\usepackage{url}
\newif\iflatexml\latexmlfalse
\usepackage[utf8]{inputenc}
\usepackage[english]{babel}
\usepackage{xcolor,listings}
\usepackage{textcomp}
\usepackage[T1]{fontenc}
\usepackage{ae,aecompl}
\usepackage{lmodern}

\usepackage{multicol}

\definecolor{myred}{rgb}{0.8, 0.0, 0.0}
\definecolor{mygreen}{rgb}{0, 0.6, 0.0}

\usepackage{indentfirst}
\usepackage{pdfpages}
\usepackage{pseudocode}

\usepackage{caption}

\def\aj{AJ}                   
\def\araa{ARA\&A}             
\def\apj{ApJ}                 
\def\apjl{ApJ}                
\def\apjs{ApJS}               
\def\ao{Appl.~Opt.}           
\def\aap{A\&A}                
\def\mnras{MNRAS}             
\def\pasa{PASA}               
\def\nat{Nature}              
\def\aplett{Astrophys.~Lett.} 

\usepackage{array}

\usepackage{enumitem}
\author[Venkatraman Krishnan et al.]{\parbox{\textwidth}{V. Venkatraman Krishnan$^{1,2,3}$, C. Flynn$^{1,2}$, W. Farah$^{1}$, A. Jameson$^{1,2}$, M. Bailes$^{1,2,4}$, S. Os{\l}owski$^{1}$, T. Bateman$^{5}$, V. Gupta$^{1}$, W. van Straten$^{6}$, E. F. Keane$^{2,7}$, E. D. Barr$^{3}$, S. Bhandari$^{1,2,8}$, M. Caleb$^{2,9}$, D. Campbell-Wilson$^{5}$, C. K. Day$^{1}$, A. Deller$^{1,2,4}$, A. J. Green$^{5}$, R. Hunstead$^{5}$, F. Jankowski$^{9,1,2}$, M. E. Lower$^{1,8}$, A. Parthasarathy$^{1,2}$, K. Plant$^{1,2,10}$, D. C. Price$^{1}$,  P. A. Rosado$^{1}$, D. Temby$^{5}$}\\ \\ \\
\parbox{\textwidth}{$^1$Centre for Astrophysics and Supercomputing,
  Swinburne University of Technology, Mail H30, PO Box 218, VIC 3122,
  Australia\\
$^2$ARC Centre of Excellence for All-sky Astrophysics (CAASTRO)\\
$^3$Max-Plank-Institute f\"{u}r Radioastronomie, Auf dem H\"{u}gel 69, D-53121 Bonn, Germany\\
$^4$ARC Centre of Excellence for Gravitational Wave Discovery (OzGrav) \\
$^5$Sydney Institute for Astronomy, School of Physics A28, University
  of Sydney, NSW 2006, Australia\\
$^6$Institute for Radio Astronomy and Space Research, Auckland University of Technology\\
$^7$SKA Organisation, Jodrell Bank Observatory, SK11 9DL, UK. \\
$^8$ATNF, CSIRO Astronomy and Space Science, PO Box 76, Epping, NSW 1710, Australia\\
$^{9}$Jodrell Bank Centre for Astrophysics, School of Physics and Astronomy, The University of Manchester, Manchester M13 9PL, UK\\
$^{10}$Cahill Centre for Astronomy and Astrophysics, MC 249-17, California Institute of Technology, Pasadena, CA 91125, USA\\
}
}\title[SMIRF I]{The UTMOST Survey for Magnetars, Intermittent pulsars, RRATs and FRBs I: System description and overview}

\begin{document}
\maketitle
\begin{abstract}
We describe the ongoing ``Survey for Magnetars, Intermittent pulsars, Rotating radio transients and Fast radio bursts'' (SMIRF), performed using the newly refurbished UTMOST telescope. SMIRF repeatedly sweeps the southern Galactic plane performing real-time periodicity and single-pulse searches, and is the first survey of its kind carried out with an interferometer. SMIRF is facilitated by a robotic scheduler which is capable of fully autonomous commensal operations. We report on the SMIRF observational parameters, the data analysis methods, the survey's sensitivities to pulsars, techniques to mitigate radio frequency interference and present some early survey results. UTMOST's wide field of view permits a full sweep of the Galactic plane to be performed every fortnight, two orders of magnitude faster than previous surveys. In the six months of operations from January to June 2018, we have performed $\sim 10$ sweeps of the Galactic plane with SMIRF. Notable blind re-detections include the magnetar PSR J1622$-$4950, the RRAT PSR J0941$-$3942 and the eclipsing pulsar PSR J1748$-$2446A. We also report the discovery of a new pulsar, PSR J1705$-$54. Our follow-up of this pulsar with the UTMOST and Parkes telescopes at an average flux limit of $\leq 20$ mJy and $\leq 0.16$ mJy respectively, categorizes this as an intermittent pulsar with a high nulling fraction of $< 0.002$.   \end{abstract}


\section{Introduction}
\noindent
Since the serendipitous discovery of radio pulsars in the late 1960s, there have been about a dozen major surveys conducted with radio telescopes in pursuit of finding more (such as \citealt{LargeEtAl1968,ManchesterEtAL1978,ManchesterEtAl1996,EdwardsEtAl2001,ManchesterEtAl2001,CordesEtAl2006,KeithEtAl2010}; see the pulsar survey database\footnote{\url{http://www.jb.man.ac.uk/pulsar/surveys.html}} by \citealt{LyonEtAL2016}). While these surveys have primarily resulted in the characterisation of a population of pulsars that have highly periodic emission, they have also led to the discovery of sub-classes of pulsars whose radio emission is rather sporadic. Of the currently known $\mathrm{\sim 2500}$ radio pulsars \citep{ManchesterEtAl2005}\footnote{\url{http://www.atnf.csiro.au/people/pulsar/psrcat/}}, only a minor fraction of them have observed pulse sporadicity. This includes $\sim200$ nulling pulsars, $\sim120$ RRATs\footnote{\url{http://astro.phys.wvu.edu/rratalog/}}, 5 intermittent pulsars and 4 radio loud magnetars \citep{Backer1970,Hesse&Wielebinski1974,Ritchings1976,Biggs1992,Duncan&Thompson1992,vanLeeuwenEtAl2002,Kramer2006a,MclaughlinEtAl2006,CamiloEtAl2006,CamiloEtAl2007,WangEtAl2007,Rankin&Wright2007,HerfindalEtAl2007,Rankin&Wright2008, Herfindal2009, Burke-Spolaor&Bailes2010, LevinEtAl2010,GajjarEtAl2012,CamiloEtAl2012,LorimerEtAl2012,LyneEtAl2010,EatoughEtAl2013,OlausenEtAl2014,LyneEtAl2017, Kaspi&Beloborodov2017, JankowskiEtAl2019}. It has long been speculated \citep{MclaughlinEtAl2006,KeaneAndKramer2008,Lyne2009,LyneEtAl2017} that sporadic emitters could well outnumber regular pulsars in the Galaxy, as the selection biases in surveys hinder their detection.

Sporadicity in the observed pulse trains can arise for a number of reasons. It may be intrinsic to the pulsar, or it may be external, for instance, due to the presence of a binary companion or interstellar propagation effects. Intrinsic sporadic emission is seen in the case of nulling pulsars \citep{Backer1970,WangEtAl2007}, mode-changing pulsars, intermittent pulsars \citep{Lyne2009}, Rotating Radio Transients (RRATs; \citealt{MclaughlinEtAl2006}) and radio-loud magnetars (e.g.\ \citealt{CamiloEtAl2012}). The presence of a binary companion can occult the pulses in case of eclipsing binary pulsars such as red-backs and black-widows (e.g.\ \citealt{BilousEtAl2018,MainEtAl2018}) and pulsar-main-sequence binaries (e.g.\ \citealt{JohnstonEtAl1994,ShannonEtAl2014}). Accretion of matter from the companion can also temporarily halt radio emission, as seen in transitional millisecond pulsars \citep{ArchibaldEtAl2009,JaodandEtAl2017}. Strong lensing of pulses from the intra-binary material in black-widow pulsars can also cause unusually bright pulses to be observed such as those seen in the case of PSR B1744$-$24A \citep{BilousEtAl2018}. Finally, interstellar scintillation plays an important role for low-DM pulsars by modulating the signal strength of the pulses and making them appear sporadic, on timescales that depends on the receiver bandwidth. It is important to note that the observed pulsar intermittency is a strong function of the sensitivity of a telescope. For instance, a nulling pulsar observe with a less sensitive telescope might just appear to be scintillating or mode-changing with a more sensitive one. Nevertheless, the question remains of why such a varied population of sporadic emitters form such a minor fraction of the known pulsar population.

To understand the part played by selection biases, we review pulsar search surveys to date. We restrict ourselves to the searches of the southern Galactic plane (declination, $\delta<0$; Galactic latitude, $|b| < 5$) with the rationale that most of such sporadic emitters are non-recycled pulsars; the non-recycled pulsars are mostly concentrated in the Galactic plane; the southern sky contains the richest portion of the Galactic plane; and this portion of the plane has been searched less exhaustively than the northern counterpart. To understand the part played by selection bias we consider previous surveys performed in the southern Galactic plane: \citealt{Large&Vaughan1971,ManchesterEtAL1978,KomesaroffEtAl1973,DaviesEtAl1977,DeweyEtAl1985,Clifton1986,D'AmicoEtAl1988,ManchesterEtAl1996,JohnstonEtAl1992,ManchesterEtAl2001,KeithEtAl2010,KeaneEtAl2018}.

These surveys resulted in discoveries of $\sim 1000$ pulsars. Most of these surveys were performed with the Parkes telescope, which is a single dish instrument with a narrow field of view (FoV;  0.55 $\rm deg^2$ for the 21-cm multibeam receiver \citep{Staveley-SmithEtAl1996}; $\sim 1~\rm deg^2$ for the 70-cm receiver). Although very sensitive, this resulted in slow survey speeds, taking on average $\sim 3$ years to complete, with an average time gap of 5 years between successive surveys. Summing the observing times of all the surveys regardless of completeness, sensitivity and operating frequency, an average total observing time spent on pulsar searches at a given point in the southern Galactic plane turns out to be only $\sim 2$ hours. This very low-cadence probe of the sky is the predominant reason why we are selectively biased against the detection of sporadic emitters. 

Furthermore, none of the surveys (with the exception of SUPERB) employed real-time single pulse or periodicity searches. Consequently, pulsar discovery lagged observations, sometimes by of order years (e.g.\ processing of some surveys is ongoing, or being repeated, and pulsars are still being found in archival data, although this is partly a benefit of improvements in processing speed and search algorithms). This discovery lag significantly affects our capacity to perform confirmation observations of sporadic emitters, which are best followed up as soon as possible after the observations in which they were discovered. For instance, XTE J1810$-$197, the first magnetar found to emit pulsed radio emission, was in a radio-silent state for almost 11 years until its recent revival in late 2018 \citep{LyneEtAl2018,LowerEtAl2018ATel,GotthelfEtAl2018,DesvignesEtAl2018,DaiEtAl2019}.  

To probe the intermittent pulsar population we require rapid, regularly repeated, real-time surveys of the Galactic plane. Selected interferometers offer this capacity, as they can combine large fields-of-view and high spatial resolution. The Australian Square Kilometer Array Pathfinder telescope (ASKAP) was used to perform one such survey recently \citep{QiuEtAl2019} in a fly's-eye mode configuration where all antennas are used as independent single dish telescopes thereby offering a huge field of view at the expense of reduction in sensitivity.
Here we present such a survey undertaken with the UTMOST telescope: SMIRF, the Survey for Magnetars, Intermittent pulsar, RRATs and Fast Radio Bursts (FRBs). SMIRF is the first rapid, multi-pass, real-time Galactic plane survey with an interferometer. The survey is commensally operated with an extensive FRB search programme \citep{FarahEtAl2018} and a major pulsar monitoring programme \citep{JankowskiEtAl2019}. 

In the next section, we describe SMIRF's scheduling and the data analysis pipeline. We then describe the survey sensitivity in \S \ref{sec:sensitivity} and report early results from the survey in \S \ref{sec:surveystodate} and \S \ref{sec:redetections}, before discussing future prospects for surveys of this type and drawing our conclusions.

\section{SMIRF scheduler and search pipelines}
\label{sec:instrumentation}

\subsection{SMIRF and the UTMOST backend}
\begin{figure}
  \begin{center}
    \includegraphics[scale=0.27,trim=0mm 0mm 0mm 0mm, clip]{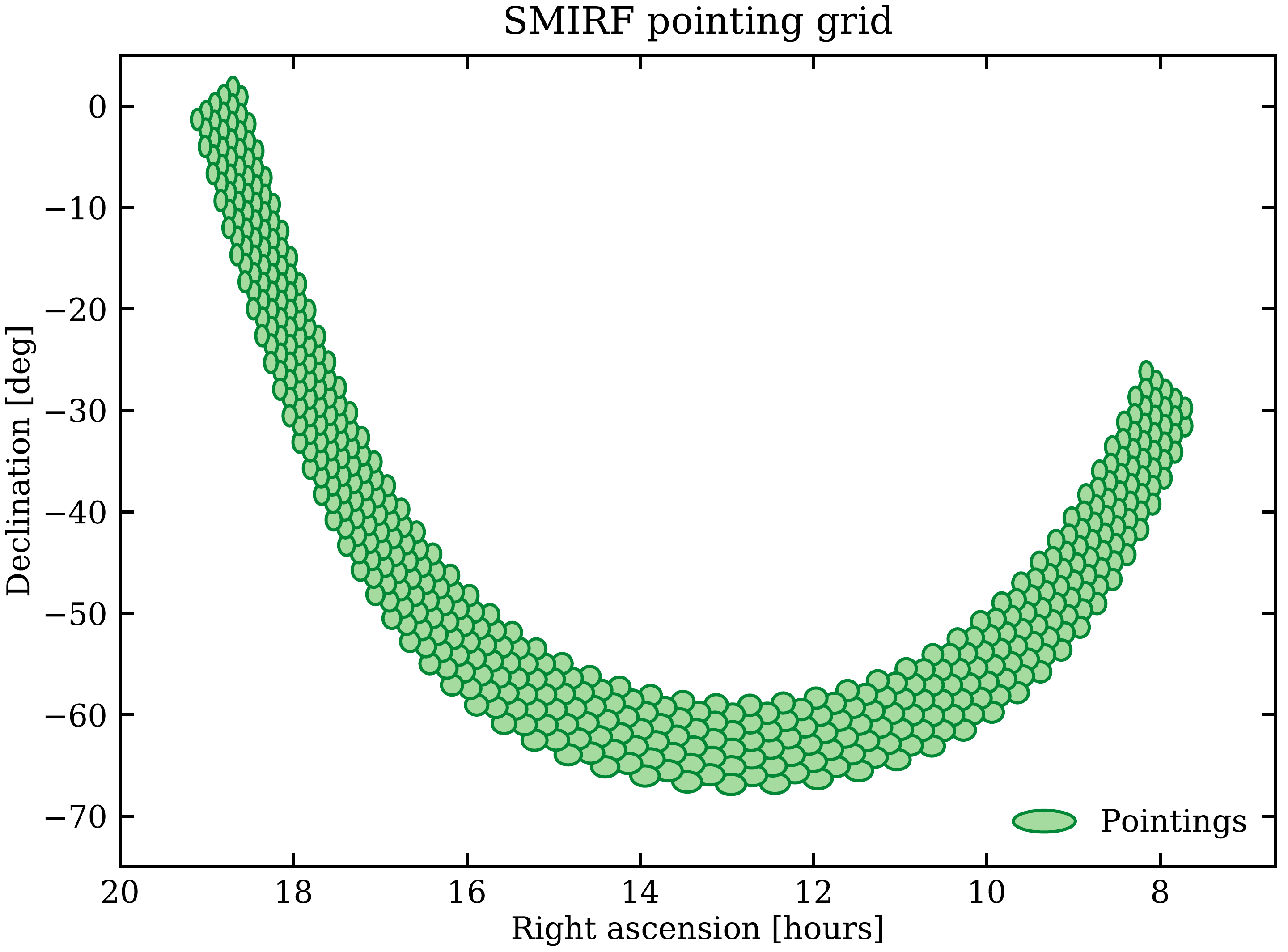}
    \caption[The grid of pointing centres for the UTMOST SMIRF survey.]{The grid of pointing centres for SMIRF, shown in equatorial coordinates. The 512 grid points are spaced on a $2\degree \times 2\degree$ grid, although the primary beam of the telescope is larger ($4\degree \times 2\degree$), so there is redundancy in the East-West direction. The extent in Galactic coordinates is  $-115\degree \leq l \leq 40\degree$  and $-4\degree \leq b \leq 4\degree$.}
    \label{fig:tile}
  \end{center}
\end{figure}

\noindent
SMIRF is a multi-pass Galactic plane survey with the capacity to search for both single pulse and periodic events in real time. The survey is conducted with the Molonglo Radio Synthesis Telescope (MOST) which is the east-west arm of the Mills-Cross telescope, 40-km South-East of Canberra, Australia. The wide field of view of the telescope enables fortnightly surveys over the southern galactic plane bound by a Galactic longitude ($l$) and latitudes ($b$): $-115\degree \leq l \leq 40\degree$  and $-4\degree \leq b \leq 4\degree$ respectively. This region is gridded into 512 pointings of $2\degr \times 2\degr$ each (see Fig.~\ref{fig:tile} for the pointing grid). A survey like SMIRF would not have been possible if not for the massively parallel computing cluster that forms the heart of the telescope's backend. We provide a brief overview of the UTMOST backend here and redirect the reader to \citet{BailesEtAl2017} for further details. Signals from 352 antennas of the 1.6-km long East-West arm of the interferometer are digitized using ``Lattice'' field programmagle gate array boards, and fed to a polyphase filter-bank (PFB) which channelizes the data from each antenna element to 320 channels with a bandwidth of $\sim100$ kHz each and a time resolution of 10.24 $\upmu$s at 8-bit resolution. The data from the PFBs are then streamed into a computing cluster via a 10-Gigabit Ethernet network, initially to a set of  ``Acquisition (AQ)'' nodes. These nodes perform phase and delay correction for each of the antennas and also perform several stages of radio frequency interference mitigation. The data are then sent via a 56 Gbps-infiniband network to a set of ``Beam Former (BF)'' computing nodes, after which all antennas for a specific frequency channel are collated into the same computing node for coherent beam forming. The coarse frequency channels are appropriately phase rotated to form 352 single-channel ``fan-beams'', laid out across the inner 4 degrees of the primary beam and equally spaced from the boresight (central) position. The single-channel fan-beams are subject to a second corner-turn where all the frequency channels for a subset of fan-beams ($\sim45$) are collated onto the same BF node. These beams are then searched in real time for transient events using a customised version of the graphics processing units (GPU) based transient detection software, \textsc{Heimdall}\footnote{https://github.com/ajameson/heimdall\_multibeam} \citep{BarsdellEtAl2012} and searched for pulsars using a new real time pulsar detection software written for this survey called ``SMIRFsoup'' (see section \ref{sec:SMIRFsoup}), which occurs within a few seconds of the observation terminating. A total of 8 BF computing nodes share the operating load of beamforming 352 fan-beams and real-time searches for single-pulses, facilitated by 3 \textsc{nvidia geforce 1080-Ti} GPUs. On average each node processes 44 fan-beams. A fourth GPU on every node is used for real-time periodicity searches, along with an additional BF node that performs the periodicity searches of ``bleeding sky positions'' (positions on the sky where data from multiple nodes need to be combined as detailed in Section \ref{sec:SMIRFsoup}).

\begin{figure*}
  \begin{center}
    \includegraphics[scale=0.40,trim=0 0 0 0, clip]{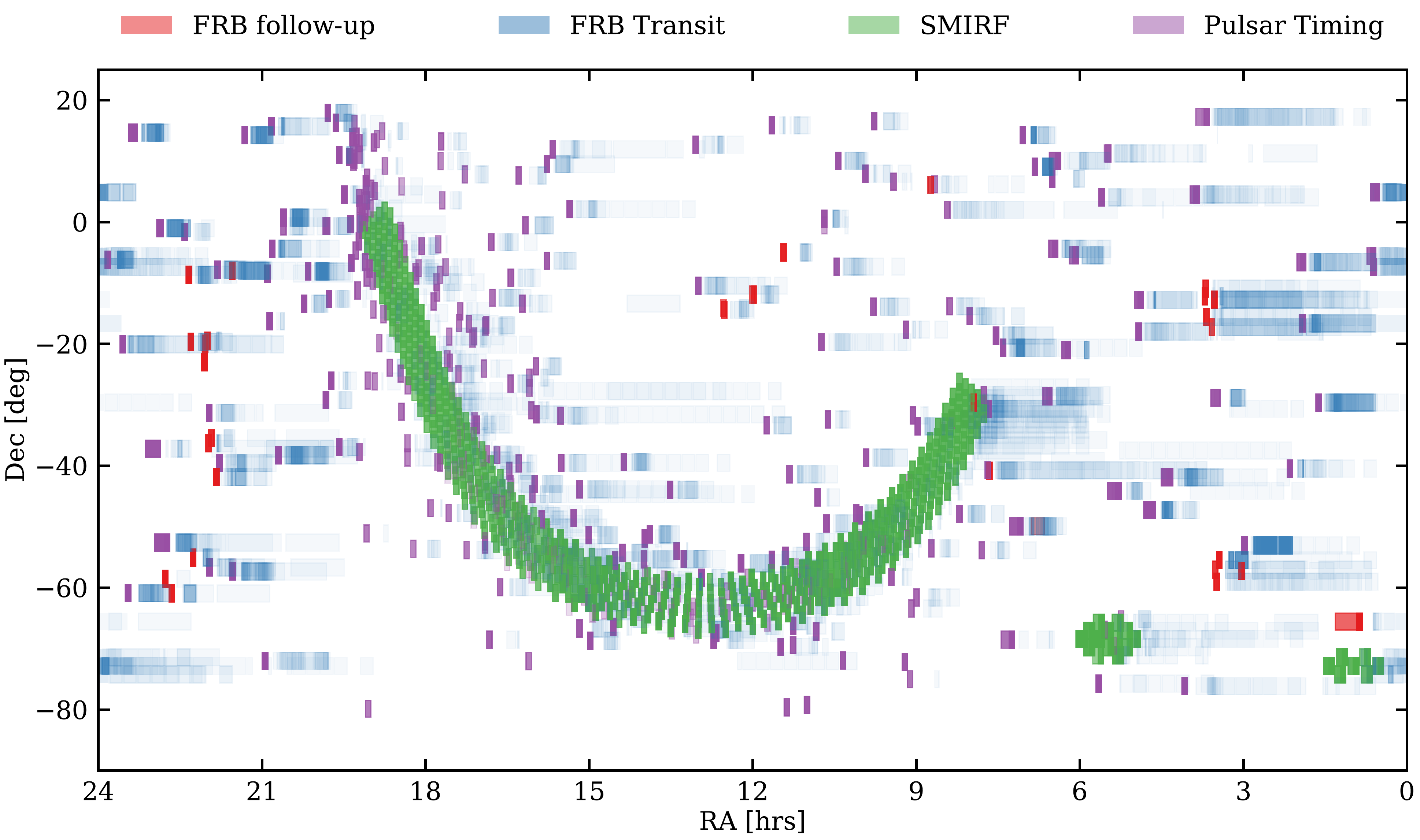}
    \caption[An example of automated and unsupervised observations of pulsars, FRB fields and SMIRF pointings by the SMIRF scheduler.]{An example of automated and unsupervised observations of pulsars, FRB fields and SMIRF pointings made by the SMIRF scheduler between Jan 2018 and June 2018. The algorithm for scheduler operations is provided in \S \ref{sec:scheduler}. The scheduler automatically monitors $\sim 500$ pulsars, $\sim 20$ FRB fields and observes $\sim 500$ SMIRF pointings with an average cadence of 2 weeks or less. Grid pointings have also been placed at the positions of the Large and Small Magellanic Clouds (green symbols, lower right).}
    \label{fig:sched_oph}
  \end{center}
\end{figure*}

\subsection{The SMIRF scheduler}
\label{sec:scheduler}
\noindent
A decision in mid-2017 to make UTMOST a transit-only instrument in the East - West direction (in order to improve its sensitivity by mitigating phasing issues and self-generated radio frequency interference) combined with a slow slew rate of the North - South drive ($\sim 5\degr~\rm min^{-1}$) meant that a sophisticated scheduler had to be developed that would maximise our observing cadence across different observing programmes. Whilst an automated dynamic scheduler developed for robotic pulsar timing observations was already in use (see \cite{JankowskiEtAl2019} \footnote{ \url{https://bitbucket.org/jankowsk/}}), it was developed for a fully steerable telescope and hence had to be extensively modified to work with SMIRF. Hence a new fully automatic scheduler was developed from scratch and added to the observing system that requires no human intervention \footnote{\url{https://github.com/vivekvenkris/SMIRFWeb}}. The scheduler automatically chooses appropriate sources/fields to observe across the observing programmes at UTMOST: namely the pulsar timing programme, the FRB search programme and the SMIRF survey itself. The pipeline performs searches for single pulses from FRBs and pulsars in real-time, periodicity searches for pulsars in real-time, commensally times up to 4 pulsars inside the primary beam of the telescope and sends email alerts of potential candidates and problems with the observing system, if any. 
The scheduler is capable of performing unsupervised $24 \times 7$ operations. Fig.~\ref{fig:sched_oph} shows the cumulative completely automated operation of the telescope from Jan 2018 to June 2018. Observations of the Galactic plane are the most prominent feature in the plot, represented by 512 ``tiles'' or pointings in green. Two similar high density observing regions are due to the Magellanic clouds, which were recently added to the survey with an observing time of 12 minutes per pointing. Remaining regions are due to off-plane pulsars, follow-up of FRBs, and various calibration sources used for phasing the array. During normal operations when there are no hardware failures in the telescope drives, the scheduler is capable of monitoring $\sim500$ pulsars, $\sim30$ FRB fields and $\sim 500$ SMIRF pointings with an average cadence of 2 weeks or less, after accounting for average maintenance downtimes. A full description of the SMIRF scheduler is given in Appendix \ref{sec:scheduler_appendix}.

\subsection{The single pulse search pipeline}
\label{sec:wael}
\noindent
SMIRF uses the single pulse pipeline that is part of the UTMOST upgrade for the FRB search programme to also perform searches for single pulses from pulsars. The pipeline is similar to the one described in \cite{CalebEtAl2016}, with the addition of a real-time classification engine \citep{FarahEtAl2018}. We provide a brief overview of its operation here. As seen in Fig.~\ref{fig:single-pulse}, the beamformed data from 352 fan-beams of an ongoing observation are streamed into a circular buffer from where they are presented to the \textsc{Heimdall} single-pulse search software. Firstly, the software incoherently dedisperses the data for a number of dispersion measure (DM) trials for a range of DMs (0--2000 pc cm$^{-3}$)\footnote{ 0-5000 pc cm$^{-3}$ from late 2018 -- to maximise FRB science)} with a tolerance of 1.25 (see Section 2.3 of \cite{LevinThesis}, and \cite{MorelloEtAl2019} for details on the tolerance parameter). The dedispersed timeseries are then convolved with a series of increasingly wide top-hat (box-car) functions with the widths ranging from 327.68 $\upmu$s to 42 ms, in steps of a factor of 2, from which potential candidates are reported. Candidates with a signal to noise ratio greater than 9 are then provided to the ``multibeam coincidencer'', which compares the DM and width of candidates detected at the same time in different fan beams to identify multi-beam detections of individual events. Any candidate that is present in more than 4 beams is classified as radio frequency interference (RFI) and is discarded from further analysis. The shortlisted candidates are then presented to a random forest machine-learning classifier \citep{Breiman2001} which classifies them as either astrophysical bursts (i.e. potential FRBs, pulses from potentially new pulsars, or pulses from known pulsars) or simply as RFI. An email alert is sent for potential FRBs and new pulsars for further follow-up. A VOEvent trigger system \citep{PetroffEtAl2017} for potential FRBs has been added recently (early 2019) for automated follow-up with partner telescopes.

\begin{figure*}
  \begin{center}
    \includegraphics[scale=0.7,trim=0mm 0mm 0mm 0mm, clip]{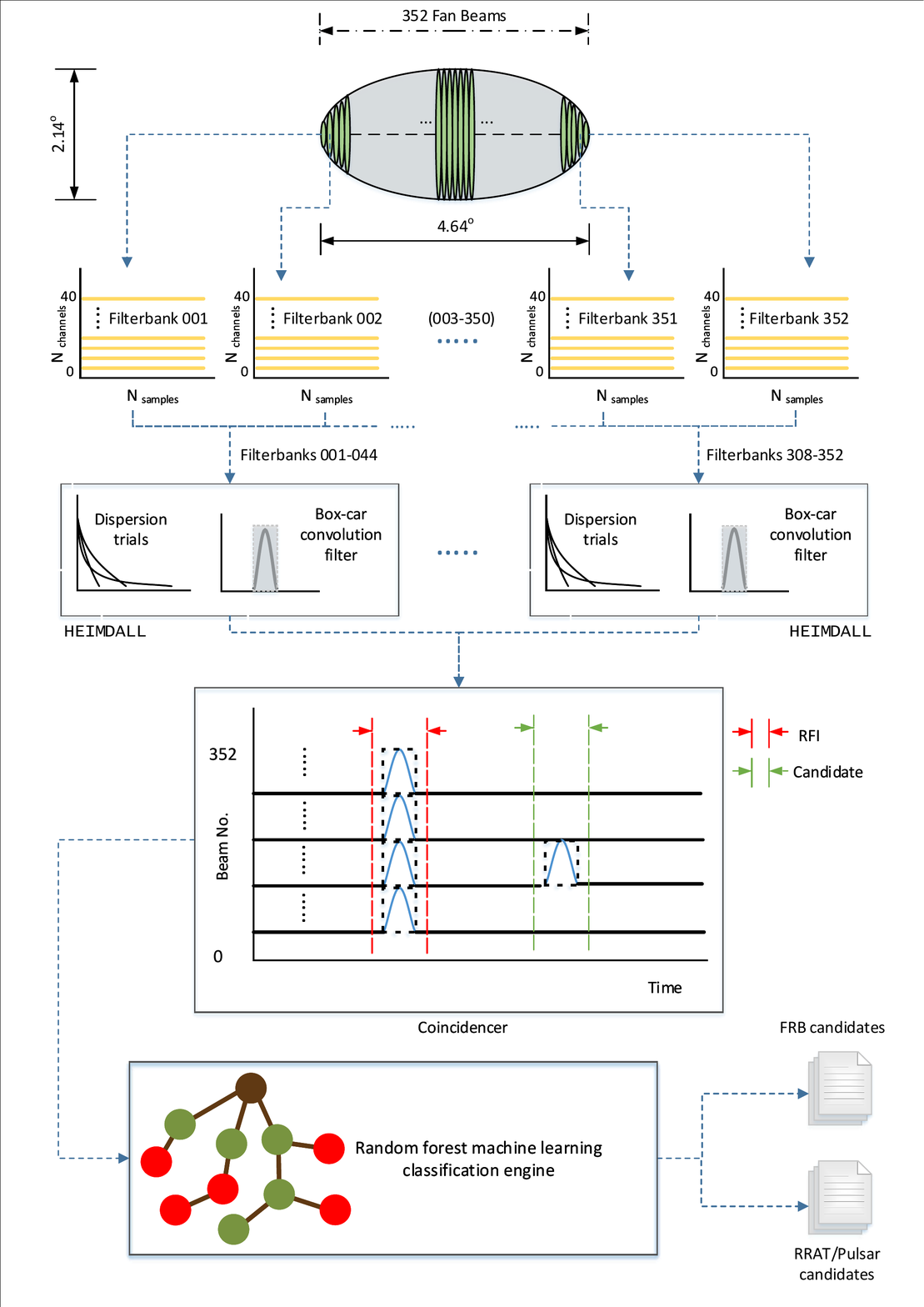}
    \caption[A schematic of the SMIRF single pulse acquisition and data analysis.]{A schematic of the SMIRF single pulse acquisition and data analysis pipeline. The primary beam is tiled with 352 fan-beams, equally spaced across 4 degrees of sky. The fan-beams produce detected time series at 327 $\upmu$sec and 97 kHz resolution, which are dedispersed in the DM range 0 to 2000 pc cm$^{-3}$ and box-car searched for single pulses down to S/N $= 9$. A random forest machine-learning algorithm classifies candidates as RFI and potential single pulses. Pulses from known pulsars are flagged and retained separately. Candidate pulses may trigger a voltage dump if they have sufficiently high DM (as part of the search for Fast Radio Bursts; see \citealt{FarahEtAl2018}).}\label{fig:single-pulse}
  \end{center}
\end{figure*}

\subsection{The SMIRF periodicity search pipeline}
\label{sec:SMIRFsoup}

\begin{figure*}
  \ContinuedFloat*
  \includegraphics[scale=0.7,trim=4 4 4 4, clip]{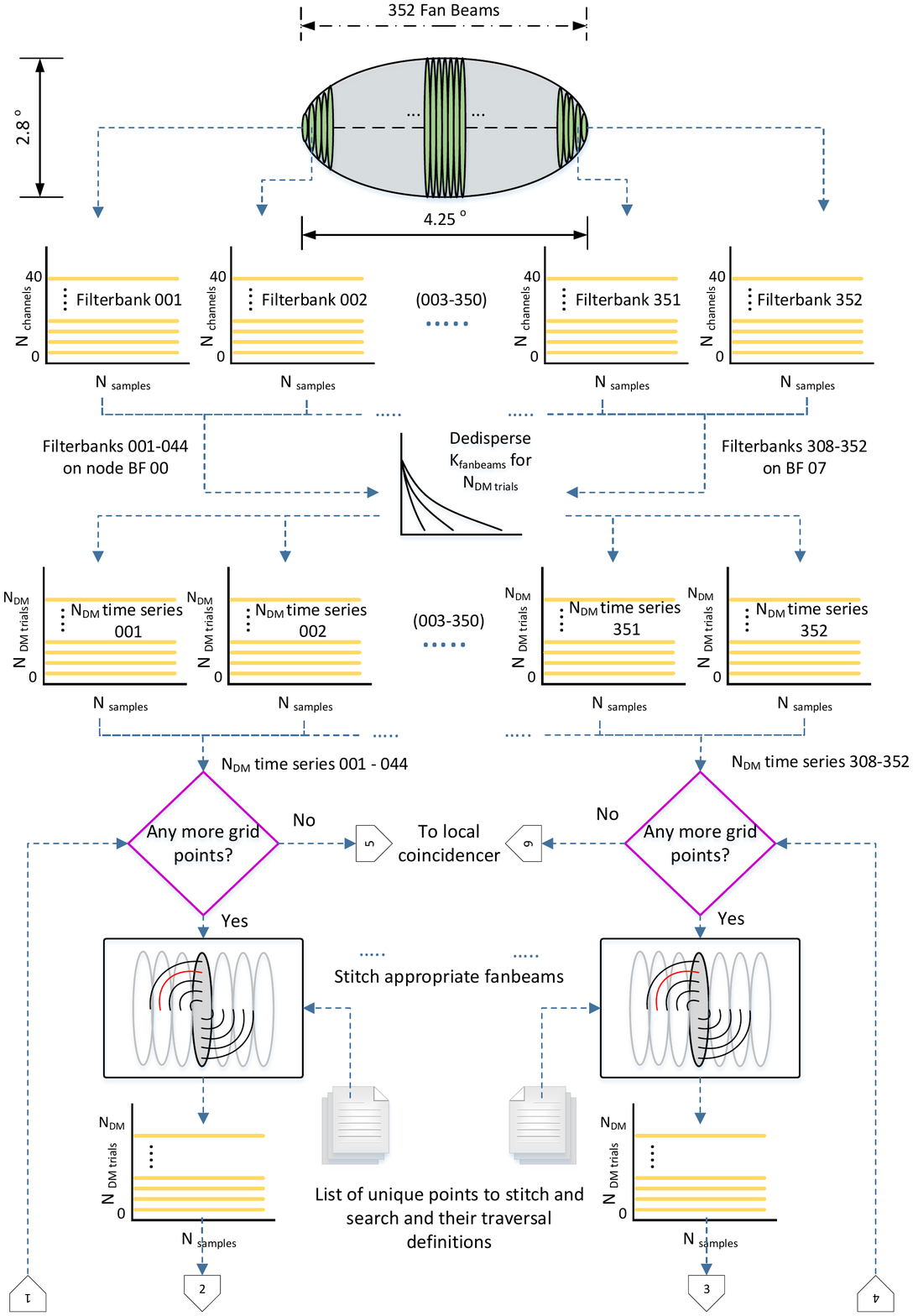}
  \caption[A schematic of the SMIRF real-time fold-mode data analysis pipeline.]{\label{fig:SMIRFsoup}A schematic of the SMIRF real-time fold-mode data analysis pipeline. As with the single pulse pipeline (see Fig. \ref{fig:single-pulse}), the data from each fan-beam is first dedispersed for $\mathrm{N_{DM}}$ trials to obtain $\mathrm{N_{DM}}$ time series. These are stitched together over a grid of tracks of points on the sky (i.e. in RA and DEC) and sent to the GPU kernel for processing. The figure is continued as Fig \ref{fig:SMIRFsoup2}} 
\end{figure*}

\begin{figure*}
  \ContinuedFloat
 \includegraphics[scale=0.7,trim=4 4 4 4, clip]{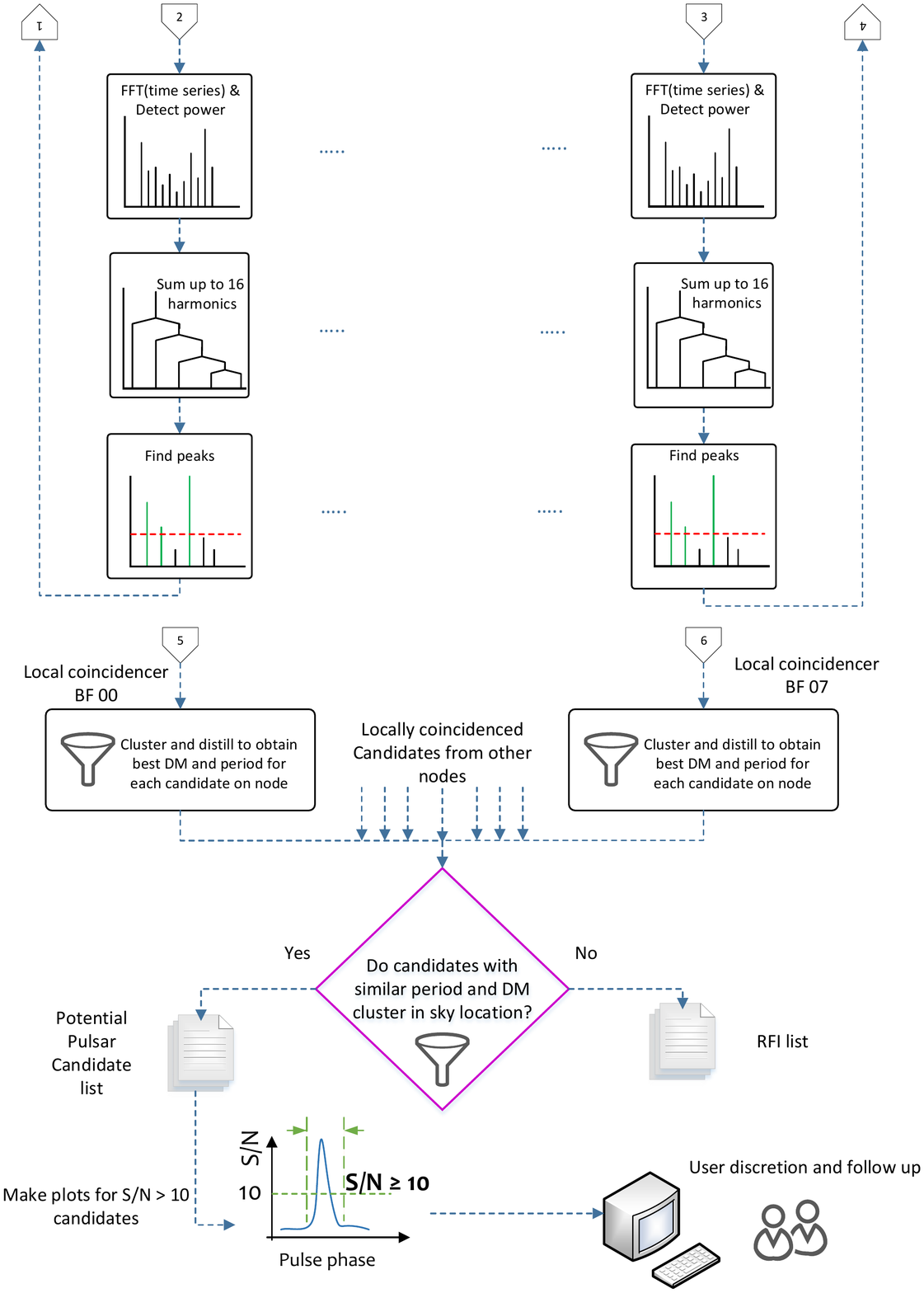}
  \caption[Continuation of Fig. \ref{fig:SMIRFsoup}.]{\label{fig:SMIRFsoup2} Continutation of Fig. \ref{fig:SMIRFsoup}. After the stitched time series arrives at the GPU, it is Fourier transformed, summed up to 16 harmonics and high signal-to-noise ratio candidates are output that are coincidenced with other DM trials and stitches to provide the final candidate list for observer discretion.} 
\end{figure*}
\noindent
To search for pulsars, we use a Fast Fourier Transform (FFT) based periodicity search software named ``\textsc{SMIRFsoup}\footnote{\url{https://github.com/vivekvenkris/SMIRFsoup}}'' which re-uses the core modules of the GPU based pulsar search software, \textsc{peasoup}\footnote{\url{https://github.com/ewanbarr/peasoup}}. \textsc{SMIRFsoup} is a heterogeneous program that performs parallel multi-threaded CPU-GPU operations to achieve real-time performance. 

Fig.~\ref{fig:SMIRFsoup} and \ref{fig:SMIRFsoup2} provide an overview of \textsc{SMIRFsoup} operations. Once an observation is complete, data from each fan-beam resides on their respective BF nodes, as \textsc{sigproc}\footnote{\url{http://sigproc.sourceforge.net/}} format filterbank files. The post-observation manager sub-thread of the scheduler (see \S \ref{sec:scheduler}) then initiates an instance of \textsc{SMIRFsoup} on every BF node and re-loads the data from all the fan-beams ($ N_{\rm beams}$) onto random access memory (RAM) of the CPU. The pipeline starts by performing an incoherent dedispersion of the data for a number of DM trials ($ N_{\rm trials}$) over the range of (0-2000 pc cm$^{-3}$) with a DM tolerance of 1.25. Once dedispersed, the data are summed in frequency to obtain ($ N_{\rm beams} \times N_{\rm trials}$) time series. 

An initial RFI rejection algorithm performs an FFT of the zero-DM time series of all the available fan-beams, and identifies periodic signals with an FFT S/N $> 9$. A local coincidencer then gathers the zero-DM ``candidates'' from every fan-beam. It shortlists the candidates that are found in more than 6 fan-beams, since, for the survey observing time per pointing, it is not possible for a pulsar to be present in more than 6 fan-beams (see below). These shortlisted candidates are then tagged as periodic RFI and are used to create a dynamic RFI-mask for the pulsar search. The identified RFI periods are also stored into an RFI database. The RFI database thus keeps a track of all the common RFI periods that have been identified by several observations.

For telescopes with a large field of view (FoV) such as UTMOST, tracking a boresight sky position of any given pointing causes a differential motion of the rest of the sky with respect to the boresight. As the field of view of UTMOST is tiled with narrow highly elliptical fan-beams, off-boresight sky positions generally move from one fan-beam to another during an observation. A potential pulsar hence might traverse several fan-beams in an observation that need to be ``stitched'' together along the path of traversal before one can use conventional FFT techniques for pulsar searches. In order to perform periodicity searches across the visible sky, we grid the visible sky into regions of ``similar'' traversals  ($N_{\rm regions}$) and model the fraction of time spent by each such region inside a given fan-beam. We compute the trajectory of each grid region for an observation and perform stitches along them to produce stitched beams ($N_{\rm stitches}$). We use telescope-specific coordinate system of North-South tilt (NS) and East-West tilt (Meridian distance; MD) to stitch fan-beams (see \citealt{BailesEtAl2017} for a brief overview of the telescope specific coordinates). 

The stitching algorithm is a core feature of SMIRF. Typically it operates with $ N_{\rm FB}=352$ fan-beams tiled across the 4-degree wide primary beam in the East-West direction. The algorithm first obtains the extent ($ E_{\rm i}$) of each fan-beam ($F^{\rm  init}_{\rm i}$) in hour angle (HA) and declination (DEC). This $E_{\rm i}$ is then gridded into a number of ``regions'' ($R_{\rm i}$) in declination, with a grid spacing of 23 arcseconds (i.e. half the fan-beam resolution). For every time section ($T_{\rm i}$ = 6 seconds; chosen on the basis of how quickly off-boresight regions move, and the computation time), the instantaneous HA of each $R_{\rm i}$ is computed, and the fan-beam $F_{\rm j}$ to which this HA corresponds is saved. This loop yields the traversal times of each $ N_{\rm FB} \times R_{\rm i}$. A secondary loop takes such trajectories and groups them into region-groups ($G_{\rm k}$) which follow sufficiently similar trajectories, in order to minimise the number of FFT operations. At the end of this procedure, we obtain a list of $G_{\rm k}$, each with their computed mean traversals for the observation such that no region ($R_{\rm i} \in G_{\rm k}$) suffers an incorrect fan-beam stitch for more than 10\% of the observing time. This limit is a trade off between real-time operation and the S/N degradation of a potential pulsar. 

The stitched beams corresponding to each $G_{\rm k}$ are then searched independently for pulsars. Since only a subset of beams are processed by every BF node, $G_{\rm k}$ that reside in boundary fan-beams tend to ``bleed'' into other nodes. The fan-beams corresponding to such $G_{\rm k}$ are transferred via the infiniband network to an additional BF node, where similar pulsar searches are performed. Thus there are a total of 9 BF nodes that share the workload of pulsar searching. Our heterogeneous CPU-GPU architecture allows for parallel stitching and pulsar searching operations. For every $G_{\rm k}$, the stitcher constructs the appropriately dedispersed  $N_{\rm DM} \times N_{\rm FB}$ time series to produce $N_{\rm DM}$ stitched time series data that are then transferred to the GPU. 

The pulsar search is independently performed for every $N_{\rm DM}$ time series. It starts with a de-reddening routine that uses a median-of-means algorithm to estimate the red noise power in each DM trial. This is followed by a $2^{21}$-point real-to-complex forward FFT. A square law detector then estimates the power spectrum and feeds it to a harmonic summing kernel. The kernel sums up to 16 harmonics (in powers of 2), subsequent to which peaks with S/N$\geq 9$ are identified as potential pulsar candidates. A harmonic distiller compares the estimated periods of the candidates and removes any candidates that are harmonics of another candidate with the fundamental period. The candidates are then passed to the multi-stage ``coincidencer''. Whilst these operations are performed on the GPU, the stitches necessary for the next $G_{\rm k}$ is performed and made available on the CPU to start the next search operation immediately upon concluding the current search. All candidates go through several stages of coincidencing before any is considered a potential pulsar. Once candidates for every DM trial are obtained, a DM-coincidencer compares the S/Ns of the candidates with the same period in every DM trial. The DM with the highest S/N for a given period is chosen and the other candidates are discarded. The DM-coincidenced candidates ($N_{\rm cands}$) are stored on the CPU RAM.

Once all the stitches for a given observation on a given BF node have been processed, that node streams its candidates to all other nodes for a global multi-beam coincidence operation. The candidate periods are first cross-checked with the RFI database for any known interference which may have seeped through the initial dynamic-RFI rejection, and are removed as appropriate. Secondly, a sky-position-coincidencer groups candidates that possess the same DM and period from multiple stitches. The sky-position of the one with the highest S/N is compared with the rest. Since it is \textit{a priori} known that the maximum number of fan-beams that a potential pulsar can move through in a given SMIRF observation is 6 (assuming sidereal rate), any candidate found in stitches whose $F^{\rm init}_{\rm i}$ are more than 6 fan-beams away are discarded as RFI. 

Candidates which pass through all the shortlists are then saved to a MySQL database (see Appendix \ref{sec:scheduler_appendix}) as potential pulsars, along with their corresponding stitching information and other meta-data. This triggers the post-observation manager to run a ``Folding'' thread, which reads the potential candidates and the raw filterbank files, and stitches the appropriate fan-beams (this time with full frequency information). The manager then loads the stitched beam into the pulsar folding software, \textsc{dspsr}\footnote{\url{http://dspsr.sourceforge.net}}\citep{vanStraten2011}, which then saves folded candidate archives at the candidate DM, period and sky position. These candidate archives and the raw filterbanks are then collated onto the SMIRF management node, where a daemon produces candidate plots using the \textsc{pdmp} program of the \textsc{psrchive} software package \citep{HotanEtAl2004}\footnote{\url{http://psrchive.sourceforge.net}}. The observer can later use a custom candidate viewer to view the candidate plots across all SMIRF surveys, along with other metadata such as the stitching information, names of known pulsars in the beam, and the tied-beam S/N of pulsars that were commensally timed. When viewing any candidate, a ``pulsar idenifier'' compares the candidate parameters to the \textsc{psrcat} database \citep{ManchesterEtAl2005} and suggests if it is a known pulsar. The raw filterbank files are decimated to a time resolution of 655.36~$\upmu \rm s$ and a frequency resolution of 0.78125 MHz at 8-bits per sample. These filterbanks are archived to magnetic tapes for re-processing if one finds a potential pulsar in a future SMIRF survey. Hence SMIRF has the potential of providing an initial estimate of the pulsar nulling fraction, just based on whether the candidate could be found by folding all observations of the particular pointing.

\section{Survey sensitivity}
\label{sec:sensitivity}

\noindent
\begin{table*}
\caption{Comparison of SMIRF survey parameters to the HTRU survey \citep{KeithEtAl2010}}, which is the most recent Galactic plane survey with the Parkes telescope.
\begin{tabular}{lllll}
\hline\hline
Parameters & SMIRF & HILAT & MEDLAT & LOWLAT \\
\hline\hline
Instrument & UTMOST & Parkes & Parkes & Parkes\\
Central frequency (MHz) &$\sim$835&$\sim$1352&$\sim$1352&$\sim$1352\\
Bandwidth (MHz) & $\sim$16&340 & 340 & 340 \\ 
$\Delta_{\rm chan}$ (kHz) & 100 &  390.625 &  390.625 &  390.625 \\
Survey region & $-115\degree \leq l \leq 40\degree$
                      & $\delta < 10\degree$ 
                      & $-120\degree \leq l \leq 30\degree$ 
                      & $-80\degree \leq l \leq 30\degree$\\
                      & $ |b|\leq 4$ & & $ |b|\leq 15$ & $ |b|\leq 3.5$ \\
number of bits &8 & 2 & 2 & 2 \\                      
$\tau_{\rm obs}(s)$  & 300 & 270 & 540 & 4300\\
$\tau_{\rm samp} (\upmu s)$ & 327.68 & 64 & 64 & 64 \\
Sensitivity limit $^\dagger$ ({\rm m}Jy) &15 $\pm$ 5&$\sim$0.18&$\sim$0.25&$\sim$0.35\\
\hline
\end{tabular}
\\$^\dagger$Assuming DM $=100$ pc cm$^{-3}$, a period of 1 sec and a 5\% pulse duty-cycle
\label{tab:survey_param_compare}
\end{table*}

\begin{figure}
 \includegraphics[scale=0.32,trim=4 4 4 4, clip]{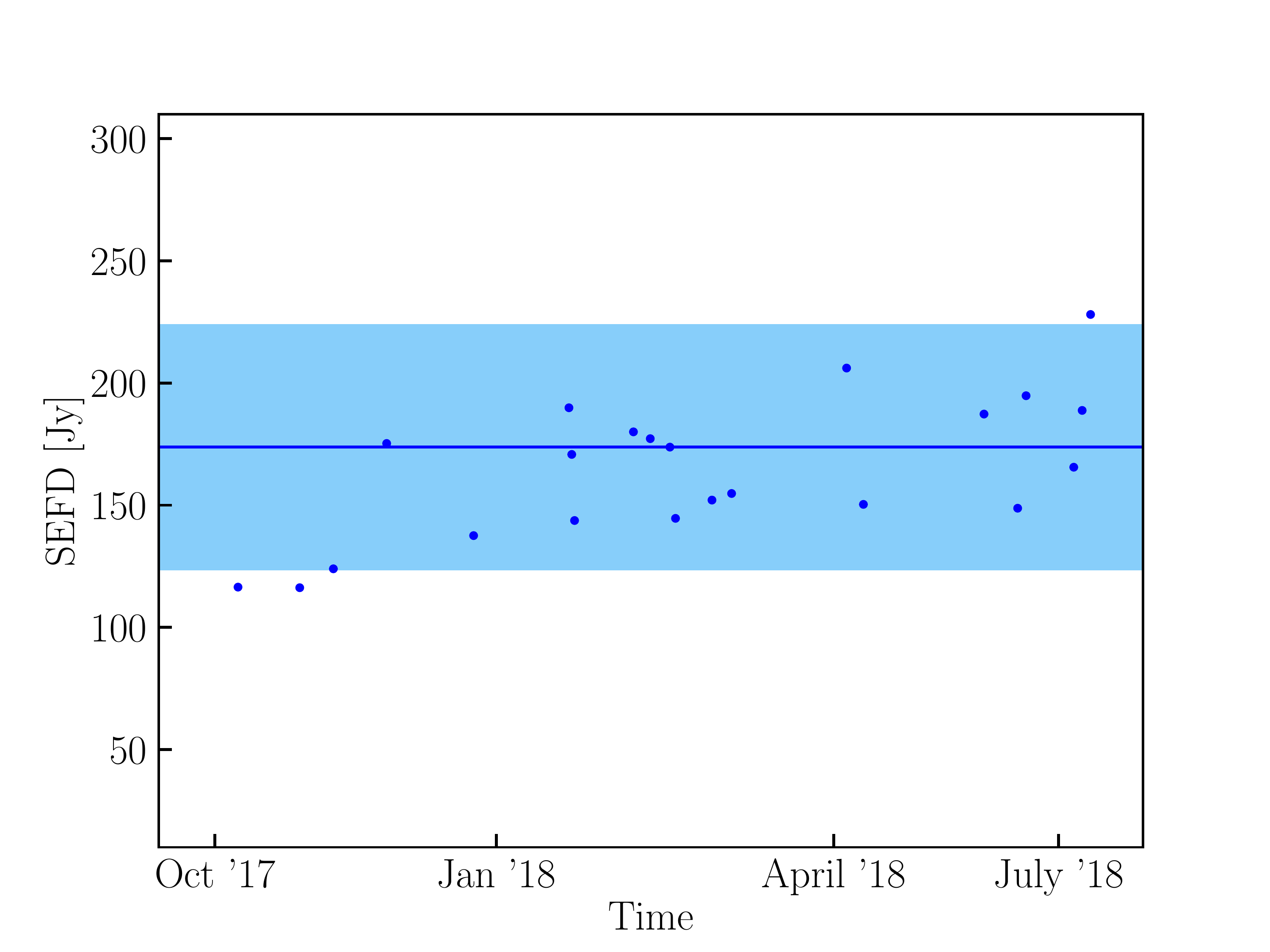}
  \caption[The estimates of System Equivalent Flux Density (SEFD) of UTMOST from observations of PSR J1644$-$4559.]{System Equivalent Flux Density (SEFD) of UTMOST derived from observations of the bright southern pulsar J1644$-$4559. The dark blue points mark the measured sensitivities and the dark blue line shows the mean SEFD of 170 Jy. The light blue band shows the 1-sigma deviation of the SEFD estimates. The RFI environment at UTMOST, which includes traffic from mobile handsets in the 820-850 MHz band, causes considerable day-to-day variation in the system sensitivity. } 
  \label{fig:survey_sefd}
\end{figure}

\begin{figure}
 \includegraphics[scale=0.32,trim=4 4 4 4, clip]{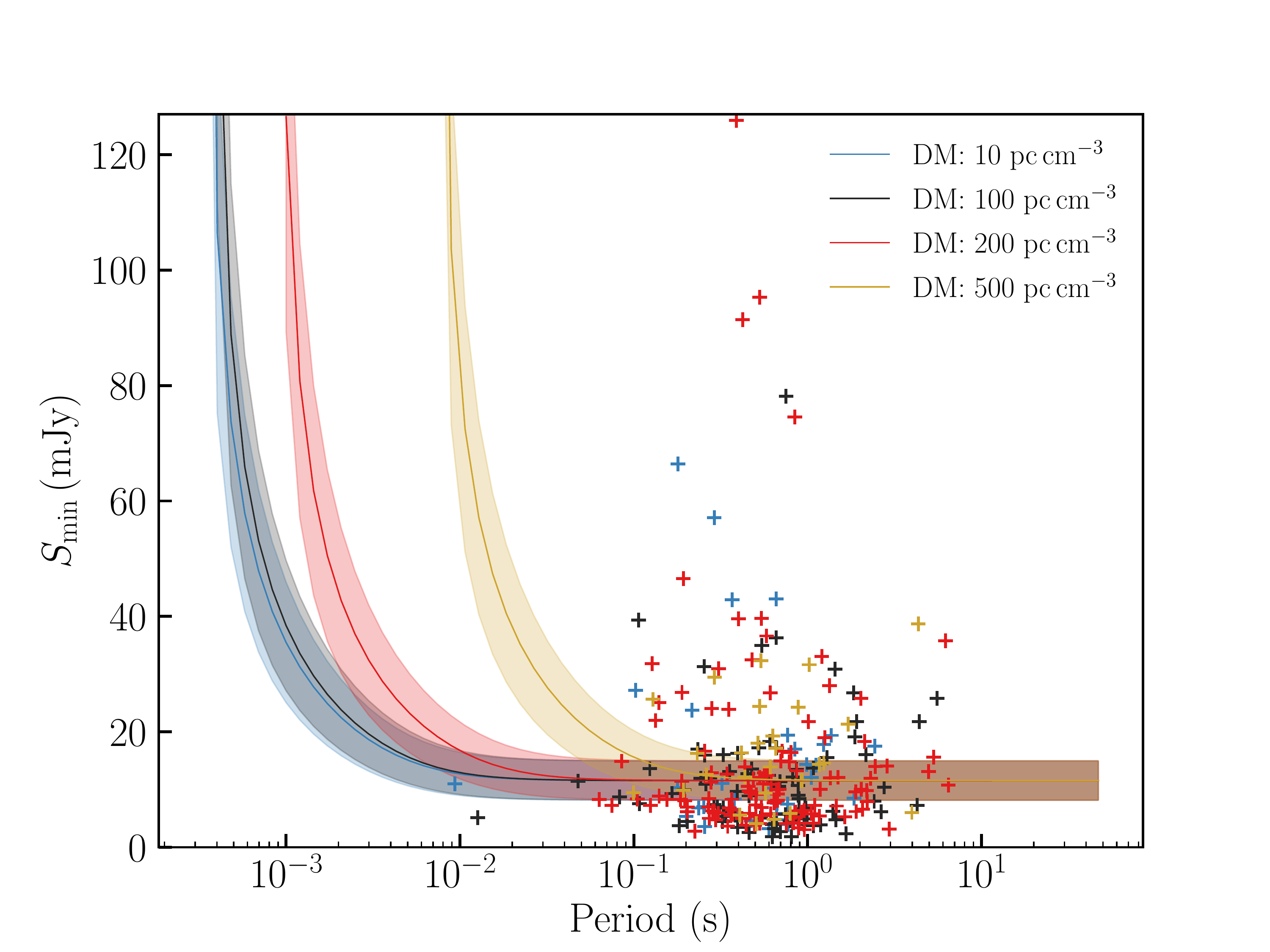}
  \caption[The mean limiting flux density for the UTMOST SMIRF survey.]{The mean limiting flux density for the UTMOST SMIRF survey shown as a function of pulsar period for various values of DM. Values shown here assume a pulse duty cycle of 5\%. The bands of the sensitivity curves are due to the uncertainty in the telescope's SEFD as shown in Fig.~\ref{fig:survey_sefd}. The crosses indicate pulsars that are part of the UTMOST timing programme and the colours indicate their corresponding DM curves. } 
  \label{fig:survey_sensitivity}
\end{figure}
The limiting sensitivity of the survey in single-pulse and fold-mode observations can be obtained from the (pulsar-specific) radiometer equation \citep{DeweyEtAl1985,Lorimer&Kramer2005} \footnote{The duty cycle term on the far right is ignored for single-pulse estimates},

\begin{equation}
\label{eqn:radiometer_eqn}
S_{\rm lim} = {\rm S/N} \frac{T_{\rm sys}}{G\sqrt{\Delta\nu N_{\rm p}t_{\rm obs}}} \sqrt{\frac{W}{P-W}}
\end{equation}

\noindent where ${\rm S/N}$ is the detection signal-to-noise ratio threshold, taken to be 9.0, $T_{\rm sys}$ is the system temperature, $T_{\rm sky}$ is the sky noise temperature, $G$ is the telescope gain in $K/Jy$, $\Delta\nu$ is the receiver bandwidth in Hz, $N_{\rm p}$ is the number of polarizations, $t_{\rm obs}$ is the time per observation in seconds, $P$ is the period of the pulsar in seconds and $W$ its pulse-width in seconds.

Although the system can record up to $31.25$~MHz of bandwidth, the high degree of gain variations across the bandpass effectively reduces the usable bandwidth to only about 16 MHz. The system equivalent flux density (SEFD =$ G/T_{\rm sys}$) at UTMOST is variable in time due to issues including incorrect phasing, phase degradation with time, ring antenna misalignment, self-induced RFI and cross-talk, high degree of variations in receiver box performance, and mechanical factors such as deformations in the telescope structure (see \cite{CalebEtAl2016} and \cite{BailesEtAl2017} for more information).  This makes it impossible to provide a single canonical sensitivity limit for the survey. Since the degradations from these contributors is difficult to measure realistically with the required cadence, an approximate SEFD value is thus obtained by using regular observations of a high-DM pulsar, PSR J1644$-$4559, and using the radiometer equation to obtain the SEFD of the telescope, assuming a flux density of $S_{\rm mean} = 960$~mJy for the pulsar at 843 MHz (from \citealt{JankowskiEtAl2019}).

Assuming an effective bandwidth $\Delta\nu = 16$ MHz and a system equivalent flux density of 170 Jy, we expect a redetection of $\sim140$ known pulsars at the present sensitivity, without correcting for position of the pulsar in the primary beam. Fig.~\ref{fig:survey_sensitivity} provides a plot of our sensitivity to pulsars as a function of pulse period for various values of DM for a fixed duty cycle of 5\%. The flux densities of known pulsars that are currently part of the timing programme, scaled to the position in the primary beam of the pointing, is also shown. Table \ref{tab:survey_param_compare} provides a comparison of the survey parameters to the HTRU survey.
\begin{figure}
 \includegraphics[scale=0.32,trim=4 4 4 4, clip]{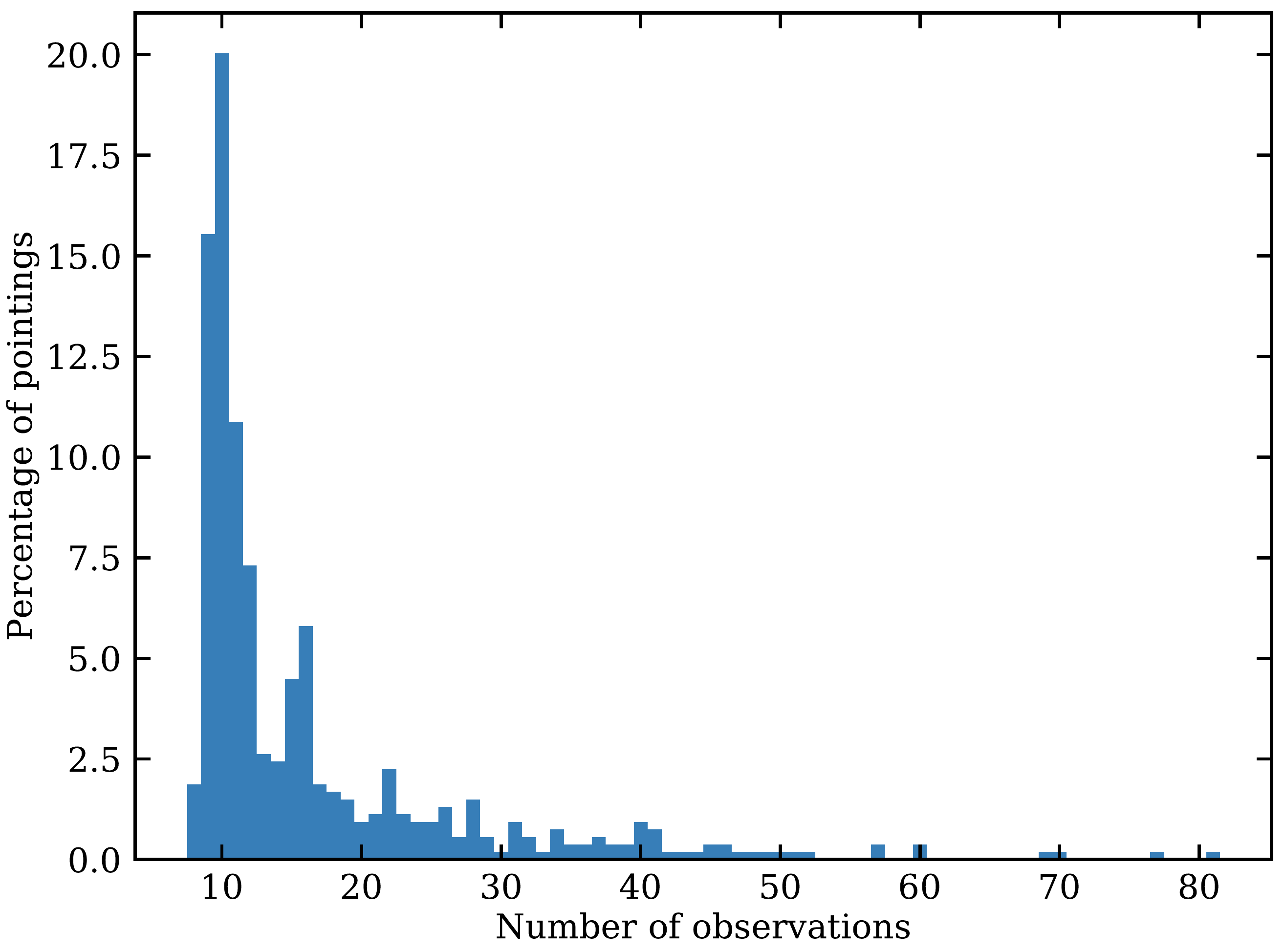}
  \caption[The fraction of survey pointings shown versus the number of times each one has been observed]{The relative fraction of survey pointings shown as a function of the number of times each has been observed. Over 98\% of the SMIRF pointings have been observed more than 10 times during 10 sweeps of the Galactic plane in the period Jan 2018 to June 2018.} 
  \label{fig:survey_hist}
\end{figure}

\begin{figure}
\centering
\begin{tabular}{cc}
 \includegraphics[scale=0.32]{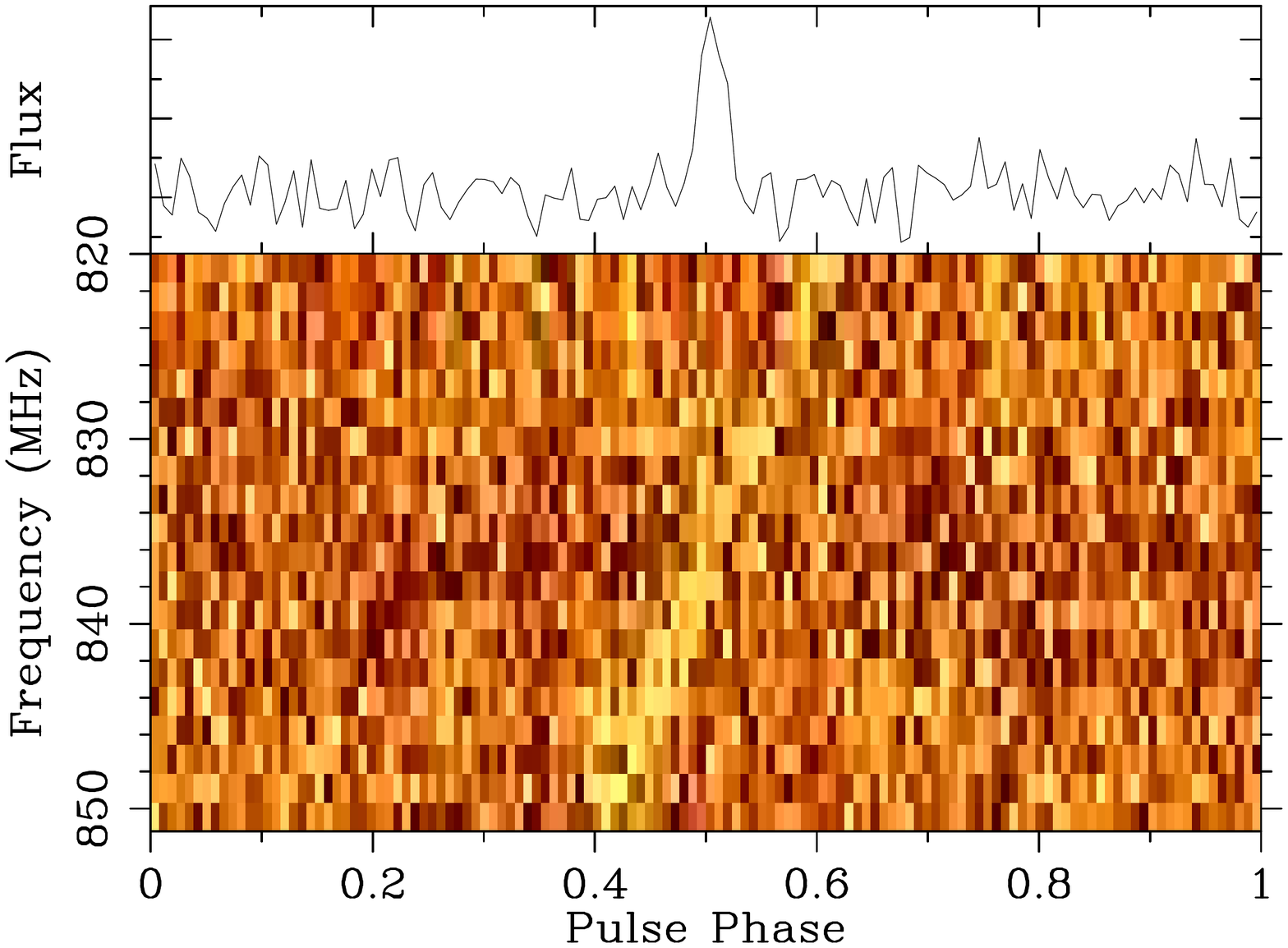} &
 \end{tabular}
  \caption[Blind detection of known pulsars as UTMOST candidates in single-pulse mode.]{An example of a blind detection of an RRAT: a single pulse detection of the RRAT PSR J0941$-$3942.} 
  \label{fig:known_detection_single_pulse}
\end{figure}

\begin{figure*}
\begin{tabular}{cc}
 \includegraphics[scale=0.3]{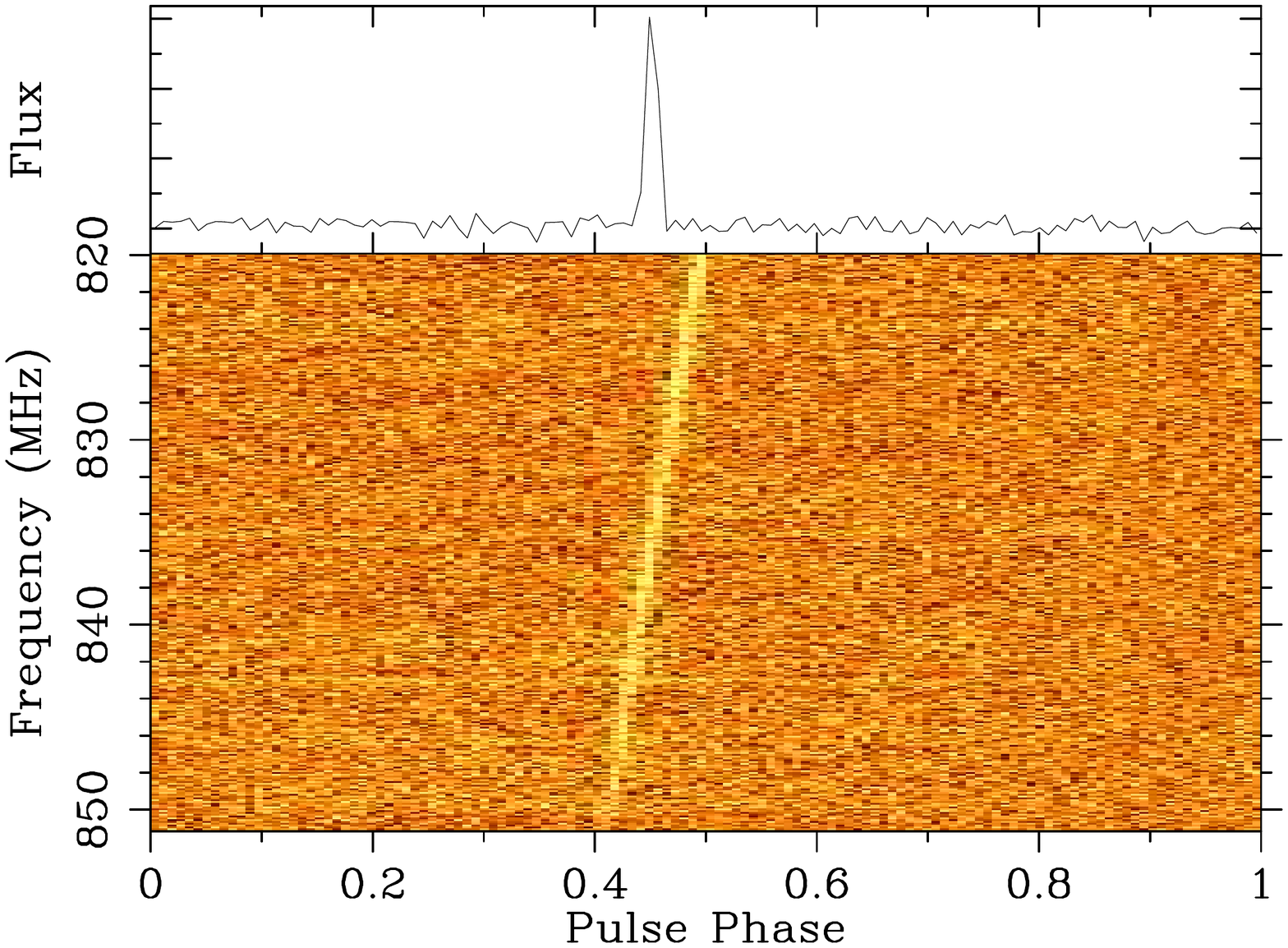} &
 \includegraphics[scale=0.3]{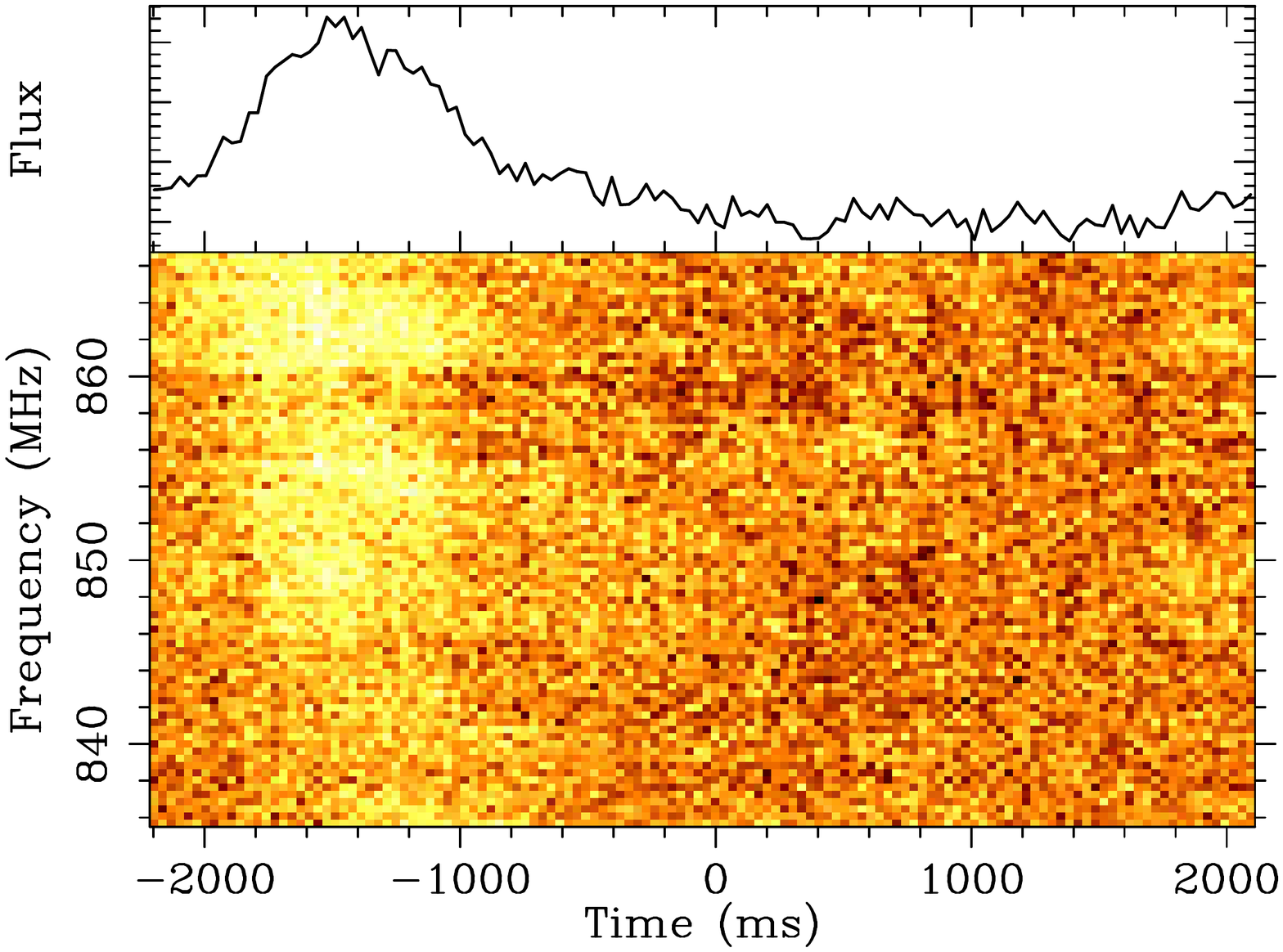}\\
 \end{tabular}
  \caption[Blind detection of known pulsars as UTMOST candidates in fold-mode.]{Example of a blind detection of known pulsars as UTMOST candidates. The left panel shows a phase versus frequency plot of the redetection of a bright pulsar, PSR J0837$-$4135. The right panel shows a phase versus frequency plot of the redetection of a magnetar, PSR J1622$-$4950, when it was caught in a radio-loud state.} 
  \label{fig:known_detection1}
\end{figure*}

\begin{figure*}
\begin{tabular}{c}
    \includegraphics[scale=0.55,trim=0 180 4 4, clip]{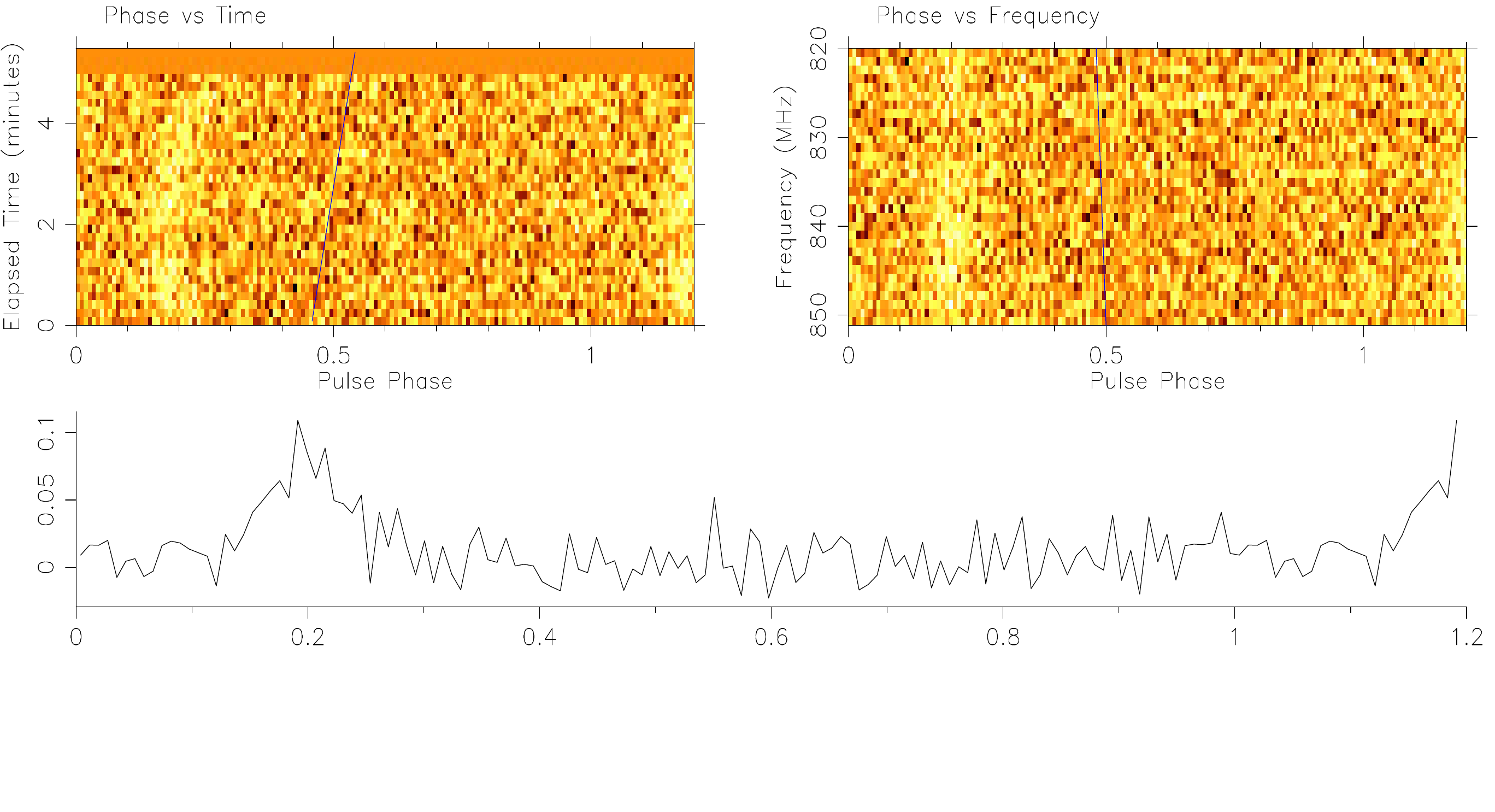}\\
 \end{tabular}
  \caption[Blind detection of known pulsars as UTMOST candidates in fold-mode.]{Phase versus frequency and time plots along with the pulse profile of the re-detection of an eclipsing binary millisecond pulsar, PSR J1748$-$2446A which resides in the Terzan 5 globular cluster (the acceleration due to the binary orbit can be seen as a bend in the pulse phase versus time panel of the plot). This pulsar is normally invisible to Molonglo's sensitivity, however, detected now from its giant pulses during eclipse egress. } 
  \label{fig:known_detection2}
\end{figure*}

\section{SMIRF surveys to date}
\label{sec:surveystodate}
\noindent

A pilot survey (SMIRF-0) was conducted to understand the real-time performance and feasibility of the survey in mid 2017, while the telescope was still capable of slewing in East-West by differential rotation of its ring antennas. Although this survey detected only $\sim30$ pulsars owing to the low sensitivity of the telescope at the time (SEFD$\sim 400 $~Jy), it was a technological milestone, as it proved we could perform a survey of the Galactic plane (gridded as Fig.~\ref{fig:tile}) in $\sim 10$ days while searching for pulsars and FRBs in real-time.

Soon after the telescope was made a transit-only instrument in the east-west direction, our sensitivity improved by a factor of 2 to SEFD=$170 \pm 30 $~Jy. After the development and testing of the fully automated scheduler, the SMIRF survey has been undertaken almost daily since Jan 2018, apart from downtime due to telescope maintenance and/or repair. At present the scheduler regularly times $\sim 500$ pulsars with a cadence of two weeks or better. This has resulted in the detection of a recent rotational-glitch in the pulsar J1709$-$4429 \citep{Lower2018} to add to the 8 previous glitch discoveries (Jankowski et al. submitted), proving the capabilities for such a massive scale pulsar monitoring programme. A new FRB was also discovered (FRB20180528; Atel: \#11675) commensally during a ``wait time'' FRB search observation that was automatically scheduled while waiting for PSR J0702$-$4956 to enter the beam.The FRB was discovered with the pipeline described in \S \ref{sec:wael} (see Farah et al. in prep., for complete details). 

In the 6 months of operations from January to June 2018, we have performed $\sim 10$ surveys of the Galactic plane with SMIRF. An important caveat is that, as one can see from Fig.~\ref{fig:survey_hist}, the survey completeness is not uniform. There are a number of reasons for this. Firstly, UTMOST is a transit telescope, and the survey time is constrained by the variable sky drift rate as a function of declination ($\delta$). This causes parts of the Galactic plane with $\delta > -30 \degr$ to be less frequently probed. Secondly, the desire/strategy to observe some of the pulsars at crucial local sidereal times at high cadence, reduces the number of surveys that can be performed around the RA of those pulsars. Some pulsars, such as the magnetar PSR J1622$-$4950, and the flux calibrator J1644$-$4559 are observed across the FWHM of the telescope owing to the varied science goals, which also reduces survey time near the RA of those sources.

We provide the initial results of the re-detection of known pulsars and a new intermittent pulsar in the next sections.

\section{Initial results}

\label{sec:redetections}
\noindent
The survey has resulted in the re-detections of 51 pulsars in single-pulse mode and at least 137 pulsars in fold-mode, which is $\approx 98$\% of the pulsars we expected to find, based on the system sensitivity, the integration times per pointing, and the position of the pulsar in the beam. 

Three of these were particulary interesting: firstly, the redection of the magnetar PSR J1622$-$4950, during its revival as a radio pulsar in late 2017 (Fig.~\ref{fig:known_detection1}), secondly, the redetection of an RRAT, PSR J0941$-$3942, during a ``wait time'' single pulse search observation (Fig.~\ref{fig:known_detection_single_pulse}), and finally, a redetection of PSR J1748$-$2446A, an eclipsing millisecond pulsar in the Terzan 5 globular cluster (Fig.~\ref{fig:known_detection2}), via giant pulses emitted during egress.

\begin{table}
\caption{Parameters for PSR J1705$-$54}
\begin{tabular}{ll}
\hline\hline
\multicolumn{2}{c}{Best-fit parameters} \\
\hline
Pulsar name\dotfill & J1705$-$54 \\
Pulse frequency, $\nu$ (s$^{-1}$)\dotfill & 1.8363(5) \\
Right ascension, $\alpha$ (hh:mm:ss)\dotfill & 17:05:37.7(0.01) \\
Declination, $\delta$ (dd:mm:ss)\dotfill & $-$54(3) \\
Dispersion measure, DM (pc~cm$^{-3}$)\dotfill & 134(15) \\
Epoch (MJD)\dotfill & 58196.8 \\
\hline
\end{tabular}
\label{tab:1705}
\end{table}

\begin{figure*}
\begin{tabular}{cc}
 \includegraphics[scale=0.30]{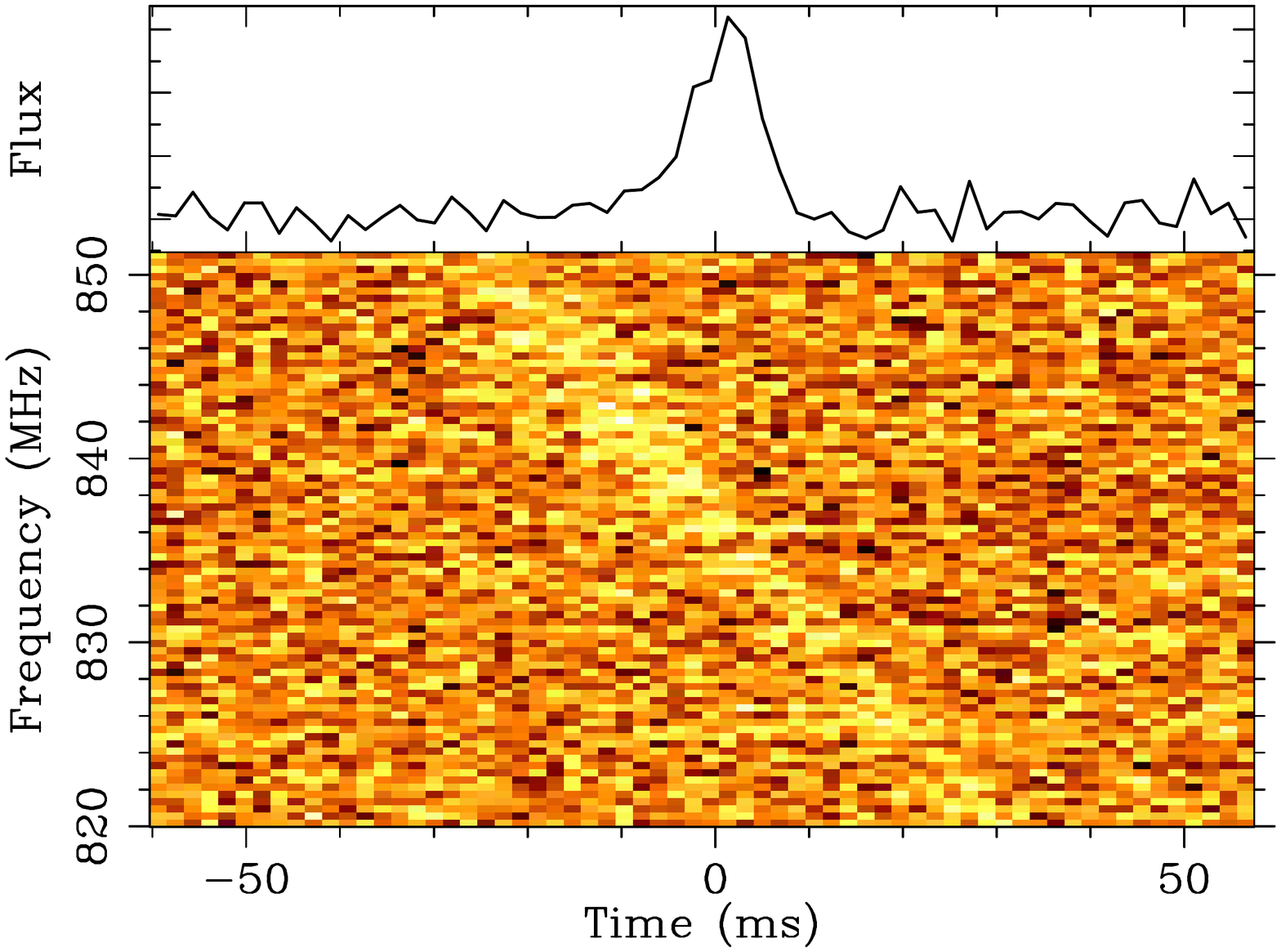} &
 \includegraphics[scale=0.30]{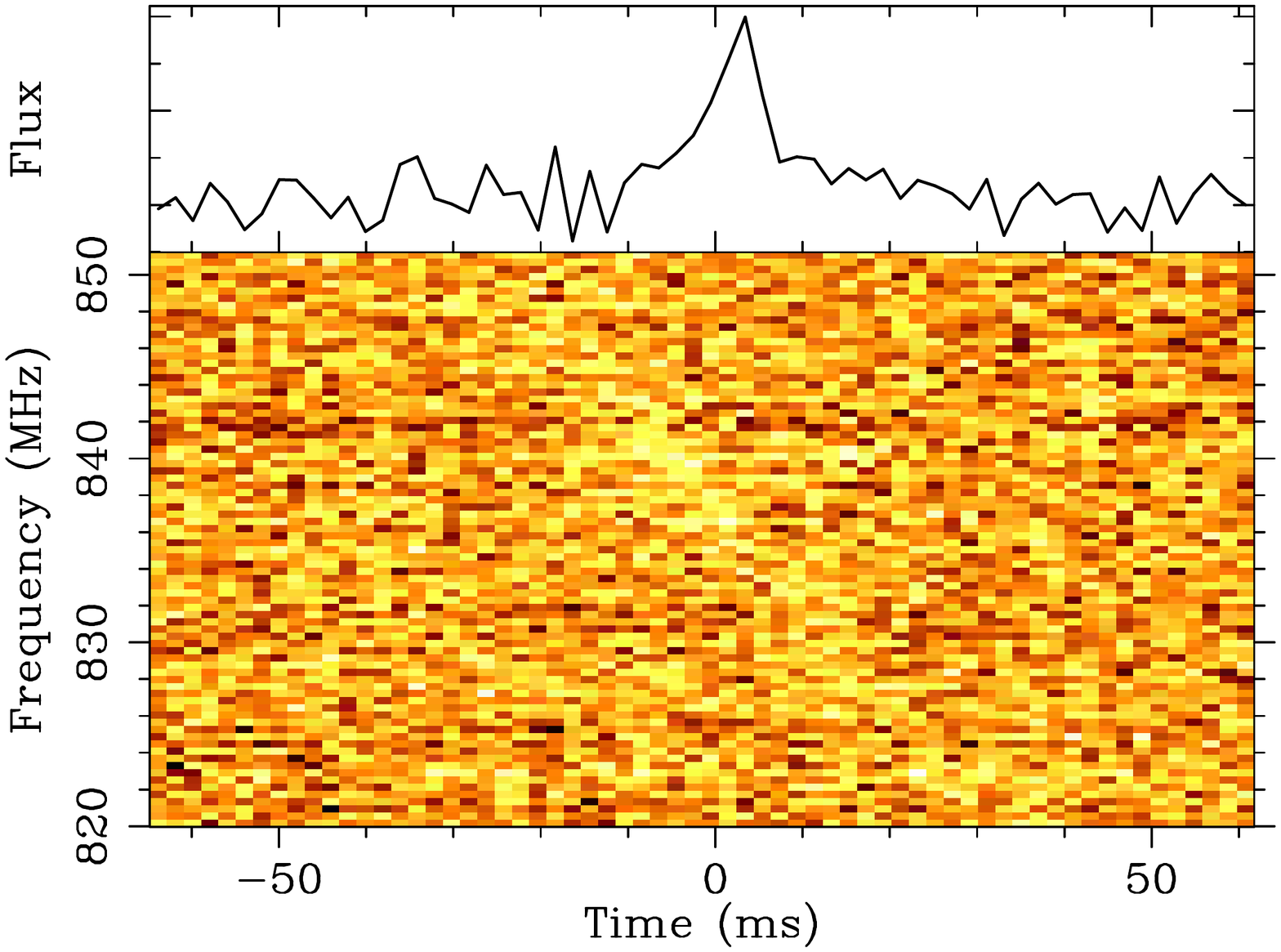} \\
 \end{tabular}
  \caption{Examples of single pulses detected from the new intermittent pulsar, PSR J1705$-$54. The top section of both the panel shows the single pulse integrated profile while the bottom section shows the (dispersed) pulse as a function of time.} 
  \label{fig:single_pulse_J1705_1}
\end{figure*}

\begin{figure*}
\begin{tabular}{cc}
 \includegraphics[scale=0.6]{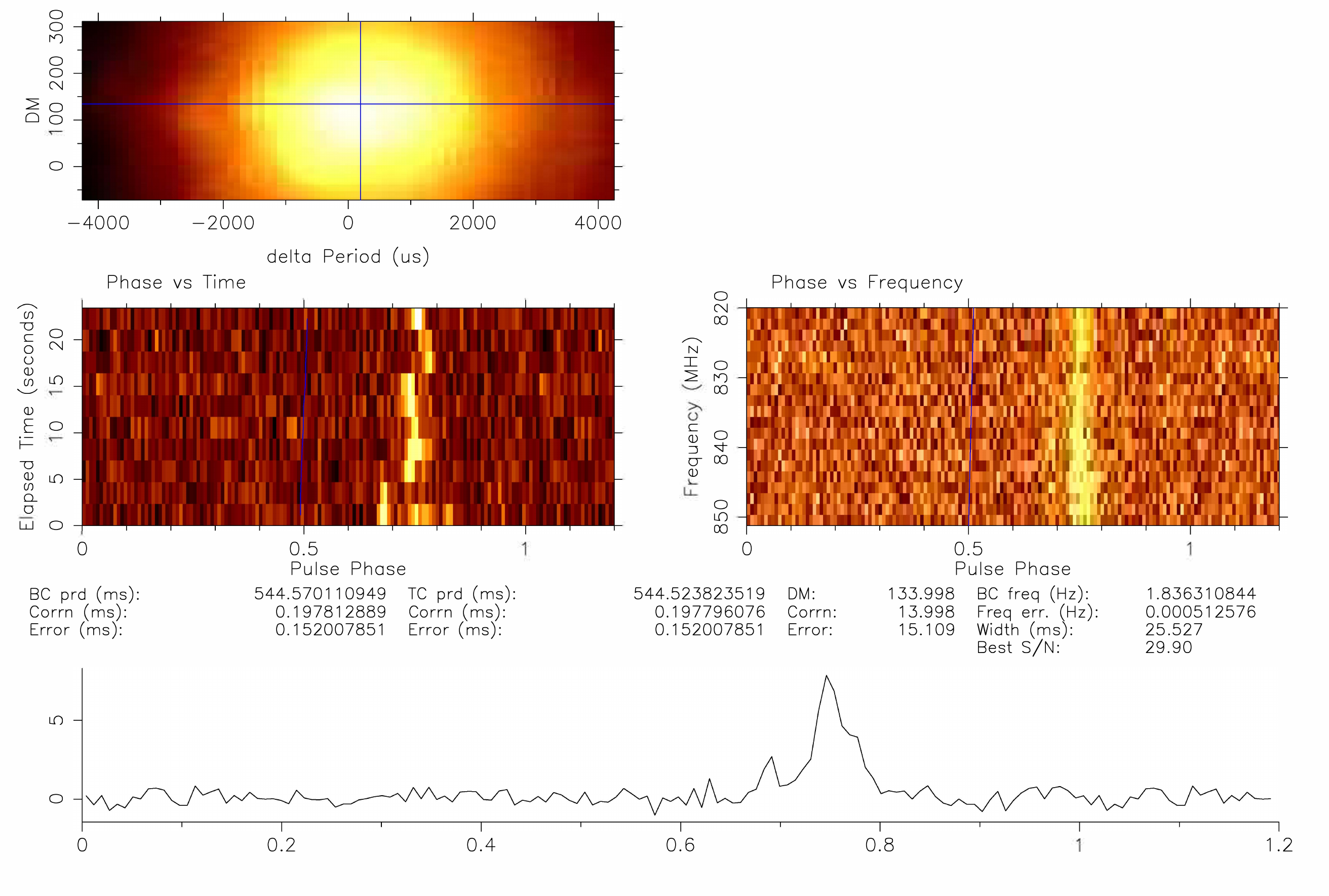}\\
 \end{tabular}
  \caption[A plot of PSR J1705$-$54 folded at the best period with single pulse time bins.]{ Detection plot of PSR J1705$-$54 folded with single-pulse time integrations. The top left panel shows the signal-to-noise ratio as a function of deviation from best DM and period. The middle panel shows the profile of the pulsar as a function of time and frequency. The bottom panel shows the integrated pulse profile of the pulsar. }
  \label{fig:combined_J1705}
\end{figure*}

\subsection{A new intermittent pulsar, PSR J1705\,--\,54}
\label{sec:1705}
As discussed earlier, the SMIRF scheduler regularly times $\sim$500 pulsars in addition to performing 2-week Galactic plane surveys. In such pointings, real-time single-pulse searches for pulsars and FRBs happen simultaneously. In one such 6-minute observation of PSR J$1711-5350$, we obtained real-time detection of single pulses from a source that was drifting across the fan-beams at the sidereal rate. Since this pointing was outside the SMIRF pointing grid, real-time pulsar searches were not performed. However, we performed an offline pulsar search and discovered a new pulsar, PSR J$1711-54$.

The pulsar was seen to be nulling during the detection observation. Out of the 6 minute observing time of the field, the pulsar was seen to be ``ON'' for only $\sim$20 seconds, out of which we obtained 8 single-pulses whose S/N was greater than our detection threshold (see Fig.~\ref{fig:single_pulse_J1705_1}  for examples of single pulses). The best-fit pulsar parameters are provided in Table \ref{tab:1705}.

Follow up observations were conducted soon after the discovery with UTMOST with real-time single pulse and fold-mode searches. 19 observations of 6 minutes each, were performed, all of which have resulted in the non-detection of the pulsar to a single-pulse and fold-mode S/N threshold of 9. This places a initial estimate of the nulling fraction of this pulsar to be $\textless 0.002$ at an average flux limit of $\sim15$ mJy. We also performed a gridded search of the sky position with Parkes as part of the SUPERB survey \citep{KeaneEtAl2018}. No single-pulses were detected at the dispersion measure (DM) of the pulsar with S/N $> 9$ (corresponding flux limit of $\sim 0.16~\rm mJy$) and folding the data with the best ephemeris of the pulsar has not resulted in a redetection. The pulsar is yet to be re-detected but regular follow-up observations are being scheduled with the UTMOST telescope by the SMIRF scheduler. Even without confirmation observations, we are confident that this candidate is a genuine pulsar because of it's non-zero DM, broadband emission, its movement from one fanbeam to another at the sidereal rate, and not close in frequency to any RFI detected with other beams or in the RFI database.

\section{Future prospects}

Although SMIRF is already performing well, there are a number of improvements that can be done to improve its pulsar discovery efficiency. In this section, we describe improvements to the data analysis pipeline that are being implemented or will be implemented in the near future.

The present pipeline performs periodicity searches over a constant DM range of 0-2000 $\rm pc ~cm^{-3}$ with a tolerance of 1.25. For some parts of the Galactic plane, searches to such high DM are inefficient, as electron density models for the ISM suggest a maximum DM in a given direction which is much smaller than this. Limiting the maximum DM to be searched for a given pointing, might be a better strategy to save computational resources that could otherwise be used to perform a search with, for instance, a better DM tolerance or to probe a modest acceleration range. For pointings toward e.g.\ the Magellanic clouds or globular clusters that are at known distances, it is also beneficial to perform coherent dedispersion of the filterbanks at the central DM of the galaxy or cluster respectively, thereby decreasing the dispersion smear of the pulsar. 

The presence of bright pulsars in the field of view (such as PSR J0835$-$4510 and PSR J1644$-$4559), can overwhelm the detection pipeline, by producing orders of magnitude more candidates than it can presently efficiently coincidence. Bright pulsars can appear as candidates in multiple stitches, up to several arc minutes away from the position of the pulsar. The coincidencer necessarily then treats each candidate as a potential pulsar, leading to several hundred candidates needing to be manually scrutinised. One can solve this problem by identifying the fan-beams these bright pulsars traverse through, and ignore any stitch that passes through these fan-beams for pulsar searching. While such a strategy would affect detections of new pulsars near such bright ones, the number of candidates to look at would be greatly reduced.

With the present pipeline, it is left to the observer to perform the manual task of searching through multiple epochs of a particular pointing to find out if an interesting candidate was also detected in early observations. Automating this task has the potential to identify sub-threshold candidates that were detected from previous surveys, but were ignored for manual scrutiny owing to low signal-to-noise ratios. These candidates would then act as verification of an interesting candidate, thereby potentially helping confirm a new detection.

The present pipeline only performs a search for pulsars in the Fourier domain via FFT techniques. As is well known, this is not ideal for detecting long period pulsars as the associated part of the spectrum is dominated by red noise. The technique can also struggle to detect pulsars with very narrow duty cycles due to the finite number of spectral harmonics that are incoherently summed to search for a peak. Fast Folding searches offers the possibility to overcome such loss in sensitivity, albeit more computationally intensive. Implementing a parallel FFA algorithm will help us remove the above mentioned bias in our processing pipeline. 

Whilst the above mentioned improvements are for the data analysis and candidate identification pipeline, SMIRF is currently mainly limited by the low sensitivity of the survey, which can only be addressed by improving the telescope hardware. While the searches with the East-West arm are to be continued in the near future, the North-South (NS) arm of the telescope is currently being refurbished to transform the telescope into a 2-D interferometer (UTMOST-2D), aimed at precise localisation of real-time detected FRBs. The NS arm is 1.55 km long and 11.7 m wide. The basic element of the NS arm (a ``cassette'') is a 1.4 m long section of the NS arm, fitted with 8 dual-polarised clover-leaf antennas that are beam-formed in analog to provide a primary beam that covers approximately $\sim 13 \degr \times 2 \degr$. The design has a system temperature of $\approx 70$ K and a bandwidth of 45 MHz centred at 830 MHz. The first stage of the project that is on-going (Day et al. in prep.), is aimed at installing 70 cassettes along the arm by end of 2019. Once completed, such a system will cover $\sim$100-m of the NS arm but will already possess a sensitivity comparable to the existing EW arm. 

It is interesting to speculate how much better SMIRF surveys would fare if performed with the NS arm. SMIRF in the EW arm is highly computationally expensive as pulsars traverse from one fan-beam to another within the duration of a transit observation. This requires re-gridding the field of view into several regions, computing their traversals, and independently performing periodicity searches along each traversals (stitches; see \ref{sec:SMIRFsoup}) On average, of order $N_{\rm stitches} \approx 5000$ stitches need to be performed for every observation, and hence $\approx 5000$ Fourier transform searches are performed. This complexity will be greatly reduced with the NS arm as the fan-beams with the NS arm are oriented along the EW direction, and hence points on the sky, in most cases, would stay within the same fan-beam for the entire observation. This reduces the number of FFT operations from $N_{\rm stitches}$ to just $N_{\rm FB}$ which for $\approx 1400$ EW fan-beams, yields a factor of $\approx$ 4 reduction in the computational requirements to perform the same survey. The released computational resources could then perform e.g.\ parallel FFA searches or acceleration searches.

If the entire 1.55 km long NS arm were tiled with cassettes, that would present an increase in sensitivity of about a factor of $\sim$10 compared to the present EW arm, and one could perform a survey to a sensitivity limit of 0.25 mJy. This sensitivity is comparable to the sensitivity of the HTRU-South LOLAT survey performed with the Parkes telescope, assuming a spectral index of $-2$ for pulsars. Importantly, as UTMOST-2D is a transit telescope, the huge NS FoV of  $\sim 13 \degr$ would make the entire latitudinal extent of the Galactic plane probed in the SMIRF survey ($|b| < 5 \degr$) fit in a single pointing for a given Galactic longitude. This means that a survey of the Galactic plane could be performed every day! Along the same line of thinking, one could perform a survey of the entire Southern sky every week. Also, similar to what is currently performed with the EW arm, the large FoV also offers the potential to monitor a massive number of pulsars $(> 1000)$ by performing commensal pulsar timing observations.

\section{Conclusions}
We have performed the first ever real-time multi-pass survey of the southern Galactic plane with the newly refurbished UTMOST telescope and reported the use of a fully automated unsupervised scheduler that is capable of intelligently observing sources/fields across the predominant observing programmes in place at UTMOST. The wide field of view and efficient scheduling permit sweeps of $\approx$ 1500 square degrees of the Southern Galactic plane to be performed in around 14 days or less, with the potential to explore a new phase space of pulsar intermittancy. We described the single pulse and periodicity search pipelines used in the survey and provided examples of known pulsar redetections. We find that the periodicity pipeline blindly recovered $\sim 98$\% of the known pulsars above the limiting sensitivity of the survey. The commensal operations with the FRB search programme and the pulsar timing programmes robotically scheduled, have resulted in the detection of 7 FRBs and 2 pulsar glitches respectively. 
We report the discovery of a new (possibly highly intermittent) pulsar PSR J1705$-$54. Finally, we discuss future modifications to the telescope and the data analysis pipeline that could improve the survey's efficiency.

\section*{Acknowledgments}
This research was primarily supported by the Australian Research Council Centre of Excellence for All-sky Astrophysics (CAASTRO; project number CE110001020). The Molonglo Observatory is owned and operated by the University of Sydney, with support from the School of Physics and the University. M. Bailes and S. Oslowski acknowledge the Australian Research Council grants OzGrav (CE170100004) and The Laureate fellowship (FL150100148). The Molonglo Observatory is owned and operated by the University of Sydney with support from the School of Physics and the University. The Laureate fellowship and the Swinburne University of Technology support the operations and upgrade of the UTMOST telescope.  ATD is supported by an ARC Future Fellowship grant FT150100415

\bibliography{mybibliography}

\appendix

\section{The SMIRF scheduler}
\label{sec:scheduler_appendix}
\begin{figure*}
  \begin{center}
    \includegraphics[scale=0.75,trim=4 4 4 4, clip]{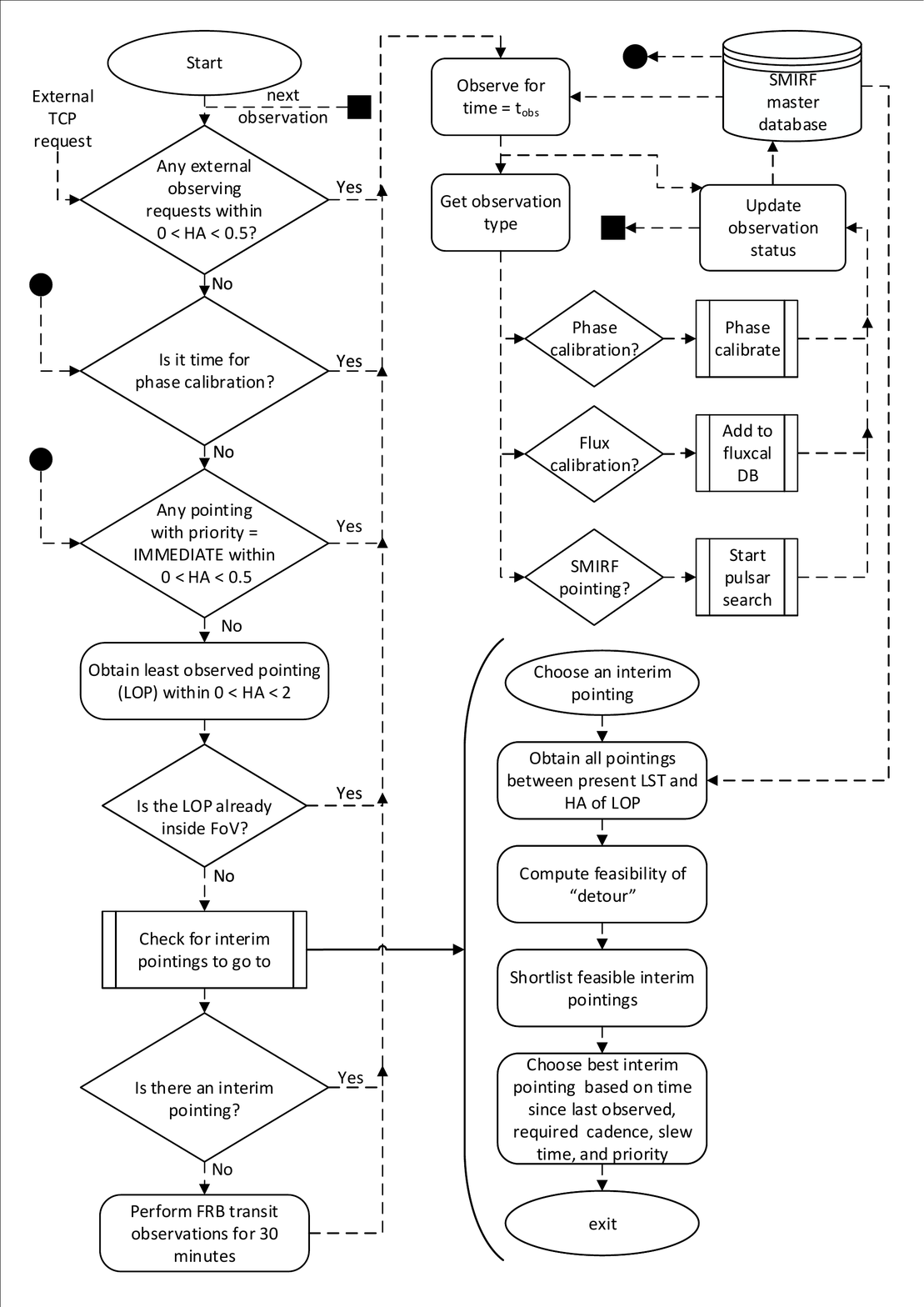}
    \caption[A schematic of the SMIRF scheduler.]{A schematic of the SMIRF scheduler. See the Appendix for details.}\label{fig:scheduler}
  \end{center}
\end{figure*}

This section describes the SMIRF scheduler that was designed to maximise our observing cadence across different observing programmes. The scheduler is capable of automatically choosing appropriate sources/fields to observe across the observing programmes at UTMOST, namely the pulsar timing programme, the FRB search programme and the SMIRF survey itself. The pipeline performs searches for single pulses from FRBs and pulsars in real-time, periodicity searches for pulsars in real-time, commensally times up to 4 pulsars inside the primary beam of the telescope and sends email alerts of potential candidates and problems with the observing system, if any. An overview of the core algorithm behind the SMIRF scheduler is detailed below. 

A MySQL backend database forms the backbone of the scheduler operations. The fields/sources of interest are first sorted into one of the following categories of pointings: a pulsar timing pointing (TIME), an FRB search pointing (FRB-SRCH), an FRB follow-up pointing (FRB-FOLLOWUP) or a SMIRF pointing (SMIRF). The pointings from TIME and FRB-FOLLOWUP which overlap within the primary beam of a SMIRF survey pointing are first removed from the list of sources to observe, except for special pointings such as our flux calibrator, a high DM pulsar, PSR J1644$-$4559. Each of the remaining pointings is firstly assigned a priority ($P^i$) to be one of $\rm \{LOW, HIGH, IMMEDIATE\}$. Each pointing also has an associated minimum dwell time($D_{ \rm min}^i$), which denotes the minimum wait time between observations before a pointing starts to ``compete'' for time again. The minimum dwell time for the SMIRF and the FRB-FOLLOWUP pointings is 4 days while the minimum dwell for the TIME pointings are decided based on their respective science cases. The millisecond pulsars J0437$-$4715 and J2241$-$5236, which are used to verify the stability of our station clock, are assigned the lowest dwell time of 2 days. The cadence for glitching and intermittent pulsars are generally higher than other pulsars. All pointings have a crucial cadence of $C\rm_{min}=20$ days after which a pointing is given the highest priority of IMMEDIATE so that it is observed soon.

The scheduler is an object-oriented and multi-threaded system designed to maximize completeness across all available pointings, mediated by the minimum cadence per pointing. Once the scheduler is started, it spawns three operational top-level threads: a status monitor thread that polls the status of the backend and the telescope position periodically, a control thread that oversees the operation, regularly monitors the priority of the pointings and sends error reports via email for any problems with the telescope operation and a core thread that handles the scheduling and observations. Fig.~\ref{fig:scheduler} provides a (highly) simplified schematic of the scheduler operation. Firstly, the scheduler checks for and chooses any user ``requested'' pointing that needs to be observed. Any pointing that ``must'' be observed on a given day, can be specified up to 24 hours in advance, via a TCP socket. Such requested pointings are assigned the highest priority. 

Secondly, the scheduler also performs automatic phase calibration observations, which can be enabled when the scheduler is initialised. If there has not been phase calibration performed in the previous 48 hours and a potential calibration quasar is about to transit the primary beam, this procedure is assigned the second highest priority. 

Thirdly, the scheduler checks for any pointing that has ``IMMEDIATE'' priority. Any pointing that has crossed $C\rm_{min}$ days since the previous observation, is dynamically provided ``IMMEDIATE'' priority by the control thread. If there are no such pointings, the scheduler then runs its the core scheduling branch.  

The branch starts by obtaining the pointing that is nearest in Hour Angle (HA) to the present telescope position which has been observed the least number of times among other ``competing'' pointings. If this least-observed pointing (LOP) is outside the primary beam of the telescope, an appropriate ``detour'' is chosen such that the telescope observes an ``interim'' pointing in the meantime. Such an interim pointing is chosen by collating all the pointings that are in between the present LST and the RA of the LOP. A first shortlist is made by choosing only the pointings which can be observed with sufficient time to get back to the declination of the LOP. Among the shortlisted pointings, the best interim pointing is chosen by performing a multi-level sorting algorithm that provides a reasonable trade-off between minimum slew-time, maximum time since the previous observation of the pointing, and its observing priority in the database. In the extremely rare case where there is no pointing to go to, an FRB search is performed at the present declination of the telescope. 

Once a pointing is chosen, the scheduler instructs the telescope to slew to the required declination. A ``wait-time'' FRB search observation is performed if the telescope is waiting for the pointing to reach a minimum hour angle, usually at the quarter-power point of the primary beam in the E-W direction. During all observations barring phase calibration, real-time searches for pulsars and FRBs are performed as detailed in the the main text. Once the observations are complete, a post-observation manager handles what needs to be done with the data. The manager calls the appropriate subroutines for phase and flux calibration observations or starts the periodicity search pipeline in case of SMIRF pointings. In case there is a problem, either with the telescope drives or the backend system, the status monitor thread will instruct the control thread to cease operations. The control thread then performs a fail-safe shutdown of several parts of the system and sends an error report as an email to the observing team.

\end{document}